\documentclass[useAMS]{mn2e}
\usepackage[tight]{subfigure}
\usepackage{longtable}
\usepackage{lscape}
\usepackage{graphicx}
\usepackage{amssymb}
\usepackage{amsmath}

\title[Mg$_{2}$ gradients in BCGs]{Mg$_{2}$ gradients as a signature of brightest cluster galaxy evolution}
\author[Loubser $\&$ S\'{a}nchez-Bl\'{a}zquez]{S. I. Loubser$^{1}$\thanks{E-mail:
Ilani.Loubser@nwu.ac.za (SIL)}, P. S\'{a}nchez-Bl\'{a}zquez$^{2}$\\
$^{1}$Centre for Space Research, North-West University, Potchefstroom 2520, South Africa\\
$^{2}$Departamento de F\'{\i}sica Te\'orica, M\'odulo C15, Universidad Aut\'onoma de Madrid, E28049, Cantoblanco, Spain}
\begin{document}


\pagerange{\pageref{firstpage}--\pageref{lastpage}} \pubyear{2011}

\maketitle

\label{firstpage}

\begin{abstract}
We have fitted the Mg$_{2}$ absorption index gradients for 21 brightest cluster galaxies (BCGs), in the nearby Universe, for which we have obtained high signal-to-noise ratio, long-slit spectra on the Gemini telescopes. This is a sub-sample of a large optical, spatially-resolved, spectroscopic sample of BCGs which allows possible connections between the kinematical, dynamical and stellar population properties to be studied. We find a weak correlation between the Mg$_{2}$ gradients and central velocity dispersion, with gradients becoming steeper with increasing mass. An equivalent correlation for normal ellipticals in the same mass range is not found, suggesting that the BCG stellar population profiles are shaped by mechanisms related to the potential well of the cluster where they live. This suggestion is reinforced by the existence of a correlation between the Mg$_{2}$ gradients and the BCG distance to the X-ray peak luminosity of the cluster.
 \end{abstract}


\section{Introduction}
The most massive of early-type galaxies -- brightest cluster galaxies (BCGs) -- are very unique, with extremely high luminosities, diffuse and extended structures, and dominant locations in clusters. Because of this special location, they are believed to be sites of very interesting evolutionary phenomena (for example dynamical friction, galactic cannibalism, cooling flows). This special class of objects may follow a separate evolutionary path from other massive early-type galaxies, one that is more influenced by their special location in the cluster. Studying BCGs may, therefore, give us information about the formation of the clusters themselves.

Numerical models based on $\Lambda$CDM standard cosmology (see, for example, De Lucia $\&$ Blaizot (2007) and Romeo et al.\ (2008) for predictions using semi-analytical techniques and full hydrodynamic simulations respectively) predict that, if cooling flows are suppressed at late times by active galactic nuclei\footnote{This assumption is necessary to reproduce the stellar populations in the centres of BCGs.}, the stars of BCGs are formed very early (50 per cent at $z \sim 5$ and 80 per cent at $z \sim 3$), and in many small galaxies. However, these models also find that BCGs assemble late: half of their final mass is typically locked up in a single galaxy after $z \sim 0.5$. BCGs, therefore, grow via accretion of small galaxies in events that do not imply new star formation.

However, not all observational findings corroborate these predictions.  For example, some studies claim that the average mass of BCGs have remained constant (or have grown very little) since $z \sim 2$ (Whiley et al.\ 2008; Stott et al.\ 2008; Collins et al.\ 2009; Stott et al.\ 2010), while other studies claim that (at least some) higher redshift BCGs are less massive than their local counterparts (Arag\'{o}n-Salamanca, Baugh $\&$ Kauffmann 1998; Collins $\&$ Mann 1998; Brough et al.\ 2002). It is still not even clear whether the evolution of BCGs depend on the properties of their host clusters (see Burke, Collins $\&$ Mann (2000) and Brough et al.\ (2002), and compare with Whiley et al.\ (2008) for contradictory results).  Despite the fact that BCGs have been extensively investigated through both observational (imaging) and theoretical studies, we do not currently have a clear picture of their formation mechanisms. Mergers between smaller galaxies, massive star formation in the early stages of the formation of the cluster due to (the now extinguished) cooling flows, or monolithic collapse that may originate from unusually large primordial fluctuations, are all possible mechanisms. These processes will leave different imprints in the dynamical properties and in the detailed chemical abundances of the stars in the BCGs. In particular, stellar population gradients derived from high signal-to-noise ratio (S/N) spectroscopy can reveal important information about the physical processes at work in galaxy formation. Such gradients and their relationships to galaxy structural parameters and the host cluster properties, therefore, provides strong constraints on galaxy formation.

Until now spectroscopic data of BCGs have been limited to either very small samples, or samples covering only the central parts of the galaxies (Tonry 1984; Tonry 1985; Fisher, Illingworth $\&$ Franx 1995a; Fisher, Franx $\&$ Illingworth 1995b; Cardiel, Gorgas $\&$ Arag\'{o}n-Salamanca 1998a; Carter, Bridges $\&$ Hau 1999; Brough et al.\ 2007; Von der Linden et al.\ 2007). This study presents a sample that is greater, in number, to all previous studies dealing with stellar population gradients in these galaxies. Furthermore, the resolution and wavelength coverage of the data allows us to apply different techniques to derive stellar population parameters (see Loubser et al. in preparation). 

This paper is part of a series of papers investigating an overall sample of 49 BCGs in the nearby Universe for which we have obtained high S/N ratio, long-slit spectra on the Gemini and WHT telescopes. The spatially resolved kinematics, central Single Stellar Population (SSP)-parameters, and the UV-upturn of the BCGs were presented in Loubser et al.\ 2008, Loubser et al.\ 2009, and Loubser $\&$ S\'{a}nchez-Bl\'{a}zquez 2010 (hereafter Paper 1, 2 and 3). A detailed study of the SSP gradients and their correlations with the host cluster properties, for a sub-sample of the BCGs with sufficient S/N ratios, will be presented in a future paper.

In the present work, we eliminate the uncertainties introduced by stellar population models, and investigate the behaviour of the Mg$_{2}$ gradients. Here, we investigate a sub-sample of BCGs with sufficient S/N ratios (described in the next paragraph), and the correlation between the Mg$_{2}$ gradients and central velocity dispersion, compared to that of elliptical galaxies in the same mass range. We also investigate the dependence of the Mg$_{2}$ gradients on the special location of these galaxies in the centre of the cluster gravitational well. Our BCG sample has the advantage of having detailed X-ray values of the host cluster properties available from the literature.

\section{Sample and data reduction}
\label{data}

Our sample comprises of the dominant galaxies closest to the X-ray peaks in the centres of clusters and, for consistency, we call these galaxies BCGs to comply with recent literature (for example De Lucia $\&$ Blaizot 2007; Von der Linden et al.\ 2007; Brough et al.\ 2007)\footnote{According to the above definition, for a small fraction of clusters the BCG might not strictly be the brightest galaxy in the cluster.}. The sample selection, spectroscopic data, and the reduction procedures were presented in Papers 1 and 2, and will not be repeated here.  The galaxy spectra of the entire sample were binned in the spatial direction to a minimum S/N of 40 per \AA{} in the H$\beta$ region of the spectrum, ensuring acceptable errors on the index measurements. Twenty-six galaxies have four or more bins when the central 0.5 arcsec, to each side, are excluded (thus in total the central 1 arcsec was excluded, comparable to the  seeing). Those galaxies that did not meet this criterion were eliminated from the final sample. Thus, the Mg$_{2}$ gradients were investigated for 25 galaxies (one was excluded because of emission contamination -- see below). The Mg$_{2}$ gradients of four of these galaxies were asymmetrical. Thus, 21 Mg$_{2}$ gradients were fitted (of which 17 were significantly different from zero, as described below).

The observations of this sub-sample were all obtained with Gemini using GMOS. The slit was placed along the major axis, although exceptions occurred if there were no suitable bright guide stars in the area in front of the mask plane, which forced the slit to be rotated to an intermediate axis (this was the case for 11 out of the 26 galaxies in Table \ref{table:objects}). On average our derived gradients reach a fraction of 0.4 of the effective radius of the galaxy (see Table \ref{table:objects} for more information on the objects). 

To use model predictions based on the Lick/IDS absorption index system (Burstein et al.\ 1984; Faber et al.\ 1985; Gorgas et al.\ 1993; Worthey et al.\ 1994), spectra were degraded to the wavelength dependent resolution of the Lick/IDS spectrograph where necessary (Worthey $\&$ Ottaviani 1997). Indices were also corrected for the broadening caused by the velocity dispersion of the galaxies (see the procedures described in Paper 2). We measured line-strength indices from the flux-calibrated spectra and calculated the index errors according to the error equations presented in Cardiel et al.\ (1998b).

We have thoroughly investigated the possible influence of scattered light on the images by interpolating across unexposed regions, and then subtracting them from the images. We have found that this contribution to the overall incident light is negligible within the spatial radii that we use here. 

We identified and corrected the spatially-resolved galaxy spectra contaminated by emission by using a combination of the \textsc{ppxf} (Cappellari $\&$ Emsellem 2004) and \textsc{gandalf} (Sarzi et al.\ 2006) routines\footnote{We make use of the corresponding \textsc{ppxf} and \textsc{gandalf idl} (Interactive Data Language) codes which can be retrieved at http:/www.leidenuniv.nl/sauron/.}, with the MILES stellar library (S\'{a}nchez-Bl\'{a}zquez et al.\ 2006c). We follow the same procedure as fully described in Paper 2. We found, and successfully removed, weak emission lines in NGC0541; NGC3311; NGC1713; NGC6173; and NGC7012. The emission in the Mg$_{2}$ region of the spectrum of ESO349-010 could not be completely removed without introducing erroneous features in the index, and therefore, the galaxy was excluded from further analysis.

The effective half-light radii $a_{\rm e}$\footnote{The effective half-light radius was calculated as $a_{\rm e} = \frac{r_{\rm e}(1-\epsilon)}{1-\epsilon \ \mid cos(\mid PA - MA \mid)\mid},$ with $\epsilon$ the ellipticity (data from NED), $r_{\rm e}$ the radius containing half the light of the galaxy (computed from the 2MASS $K$-band 20th magnitude arcsec$^{-2}$ isophotal radius as described in Paper 1), PA the slit position axis, and MA the major axis.}, and fraction of $a_{\rm e}$ which the radial profiles measured in this work spans, are listed in Table \ref{table:objects}. Gradients were obtained by a linear least-squares fit to the radial profiles, and statistical t-tests were performed on all the slopes to assess if a real slope are present or if it is zero (as a null hypothesis). For our degrees of freedom, a $t$-value larger than 1.96 already gives a probability of lower than five percent that the correlation between the two variables is by chance. $P$ is the probability of being wrong in concluding that there is a true correlation (i.e.\ the probability of falsely rejecting the null hypothesis).

\begin{table*}
\begin{scriptsize}
\begin{tabular}{l r r r c r c  r@{$\pm$}l r r r c} 
\hline Object & \multicolumn{1}{c}{PA} & \multicolumn{1}{c}{MA} & \multicolumn{1}{c}{$r_{\rm e}$} & $\epsilon$ & \multicolumn{1}{c}{$a_{\rm e}$} & Fraction of $a_{\rm e}$ & \multicolumn{2}{c}{Mg$_{2}$ Gradient} & \multicolumn{1}{c}{$t$} & \multicolumn{1}{c}{$P$}& \multicolumn{2}{c}{R$_{\rm off}$} \\
       &  \multicolumn{1}{c}{}  & \multicolumn{1}{c}{} & (arcsec) & & (arcsec) & \multicolumn{2}{c}{} & & & & (Mpc) & ref \\
\hline					
ESO146-028$\star$ & 154 & 154 & 12.4 & 0.35 & 12.4 & 0.4 & --0.069 & 0.013 & --5.31 & 0.0005& 0.051 & cb\\
ESO303-005 & 55 & -- & 9.1 & 0.25 & 9.1 &  0.4 & --0.029  & 0.012$\star$  & --2.57 & 0.0620& 0.010 & cb\\
ESO349-010 & 14 & 155 & 15.3 & 0.52 & 12.3 & 0.6 & \multicolumn{2}{c}{--} & -- & -- & --0.019 & e\\								
ESO488-027 & 88 & 68 & 10.3 & 0.15 & 10.1 & 0.4 & --0.049 & 0.004 & --11.50 & $<$0.0001& $\star$ & cb\\								
ESO552-020$\star$ & 148 & 148 & 18.3 & 0.43 & 18.3 & 0.5 & --0.028 & 0.007 & --4.06 & 0.0016& 0.013 & cb \\
GSC555700266 & 204 & -- & 10.6 & 0.30 & 10.6 & 0.2 & \multicolumn{2}{c}{--} & -- & -- & 0.020 & cb\\
IC1633$\star$ & 97 & 97 & 23.9 & 0.17 & 23.9 & 0.7 &  --0.051 & 0.003 & --17.36 & $<$0.0001& 0.015 & cb\\
IC4765 & 287 & 123 & 28.0 & 0.46 & 27.1 &  0.1 & \multicolumn{2}{c}{--} & -- & --& 0.007 & cb\\
IC5358 & 40 & 114 & 17.4 & 0.60 & 8.3 &  0.6 & --0.045 & 0.007 & --6.13 & $<$0.0001& 0.002 & cb \\
LEDA094683$\star$ & 226 & 46 & 7.3 & 0.33 & 7.3 & 0.3 & \multicolumn{2}{c}{--} & -- & -- & 0.044 & p\\
MCG-02-12-039$\star$ & 166 & 180 & 15.2 & 0.19 & 15.1 & 0.4 & --0.054 & 0.008 & --6.75 & $<$0.0001& 0.031 & e\\
NGC0533 & 350 & 50 & 23.7 & 0.39 & 18.0  & 0.2 & --0.048 & 0.013 & --3.47 & 0.0096& 0.004 & cb\\
NGC0541$\star$ & 64 & 69 & 15.1 & 0.06 & 15.1 & 0.2 & --0.015 & 0.005 & --3.04 & 0.0288& 0.037 & cb \\
NGC1399 & 222 & -- & 42.2 & 0.06 & 42.2 & 0.6 & --0.057 & 0.003 & --19.96 & $<$0.0001 & $<$0.001 & cb\\ 			
NGC1713 & 330 & 39 & 15.5 & 0.14 & 14.0 &  0.4 & --0.060 & 0.008 & --7.18 & $<$0.0001& -- & --\\
NGC2832 & 226 & 172 & 21.2 & 0.17 & 19.6 &  0.5 & --0.092 & 0.013 & --6.98 & $<$0.0001& 0.038 & cl\\ 
NGC3311 & 63 & -- & 26.6 & 0.17 & 26.6 & 0.2 & \multicolumn{2}{c}{--} & -- & --& 0.015 & pe \\									 
NGC4839$\star$ & 63 & 64 & 17.2 & 0.52 & 17.2 &0.3 & --0.035 & 0.007 & --4.84 & 0.0019& $\star$ & --\\
NGC6173$\star$ & 139 & 138 & 15.0 & 0.26 & 15.0 & 0.1 & --0.072 & 0.006 & --11.18 & 0.0079& $\star$ & --\\
NGC6269 & 306 & 80 & 14.1 & 0.20 & 13.1 & 0.3 & --0.035 & 0.007 & --4.81 & 0.0030& 0.002 & cc\\ 
NGC7012$\star$ & 289 & 100 & 15.8 & 0.44 & 15.6 &0.5 & --0.051 & 0.007 & --7.72 & $<$0.0001& -- & --  \\
PGC004072 & 204 & 83 & 10.2 & 0.33 & 8.2 & 0.2 & --0.034 & 0.014$\star$ & --2.51 & 0.1286 & 0.006 & cb\\
PGC030223 & 145 & 1 & 8.4 & 0.00 & 8.4 & 0.4 & --0.047 & 0.009 & --5.14 & 0.0021 & 0.027 & cb\\
PGC072804$\star$ & 76 & 76 & 7.6 & 0.16 & 7.6 & 0.5 & --0.057 & 0.015 & --3.67 & 0.0144 & 0.035 & cb \\ 
UGC02232 & 60 &  -- & 9.7 & 0.00 & 9.7 &  0.4 & --0.018 & 0.006$\star$ & --2.90 & 0.0442 & 0.136 & cc\\
UGC05515 & 293 & 83 & 12.0 & 0.13 & 11.8 &0.2 & --0.039 & 0.016$\star$ & --2.48 & 0.1313& 0.037 & cb \\
\hline
\end{tabular}
\end{scriptsize}
\caption{Properties of the BCGs (see Paper 1 for more detailed properties). The major axis is denoted by MA, and the slit PA is given as deg E of N. The galaxy names marked with a $\star$ is where the slit PA was placed within 15 degrees from the MA (where the MA was known). The half-light radii ($r_{\rm e}$) were calculated from the 2MASS catalogue, except ESO303-005 for which it was calculated with the \textsc{vaucoul} task in the \textsc{reduceme} package. The fraction of the effective half-light radii spanned by the radial profiles measured in this work are also listed. ESO349-010 is included in the table even though the emission lines could not be fully removed. A $\star$ denotes where the Mg$_{2}$ gradient is not bigger than three times the 1$\sigma$ error on the gradient. The five index gradients which are not indicated were excluded because the gradient profiles were asymmetric (four cases), or contained emission (one case). The $t$ and $P$ parameters of the statistical t-tests performed on the Mg$_{2}$ gradients are also shown. The projected distance between the galaxy and the cluster X-ray peak (R$_{\rm off}$) is in Mpc. The $\star$ marks at R$_{\rm off}$ indicate the galaxy is not in the centre of the cluster but closer to a local maximum X-ray density, different from the X-ray coordinates given in the literature. The references are: e = Edwards et al.\ (2007); cc = Calculated from Bohringer et al.\ (2000); cl = Calculated from Ledlow et al.\ (2003); cb = Calculated from Bohringer et al.\ (2004); p = Patel et al.\ (2006); pe = Peres et al.\ (1998).}
\label{table:objects}
\end{table*}

The central velocity dispersion values of the galaxies were derived for regions with the size of $a_{\rm e}/8$, and with the galaxy centres defined as the luminosity peaks. The index values at each radius were measured using the kinematic profiles presented in Paper 1, and the luminosity-weighted centres of the spatial bins were used to fit the Mg$_{2}$ gradients in this paper.

\section{Mg$_{2}$ gradients: results and discussion}
\label{indices}

\begin{figure*}
   \centering
   \mbox{\subfigure[ESO146-028]{\includegraphics[width=4.8cm,height=4.8cm]{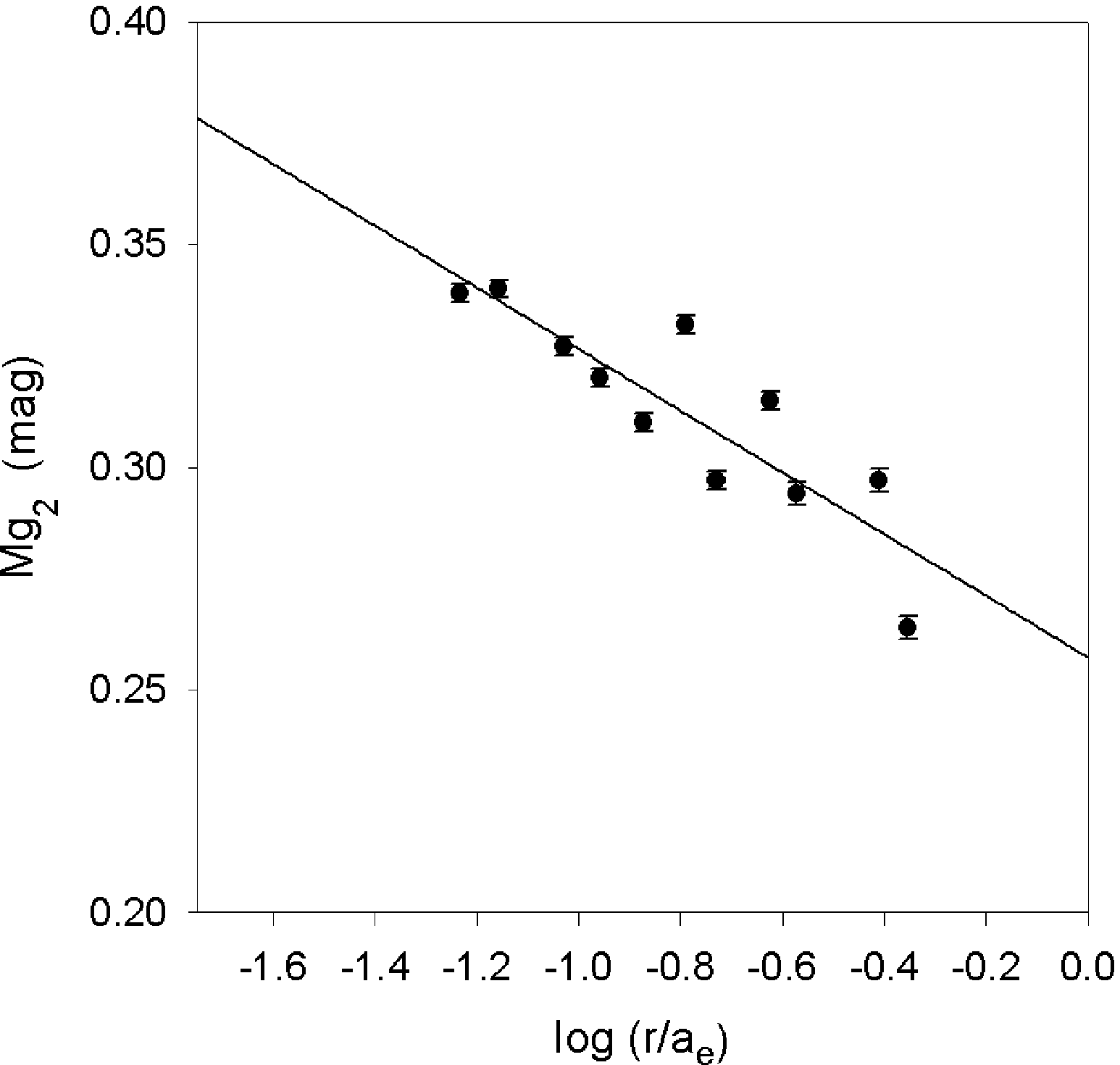}}\quad
         \subfigure[ESO303-005]{\includegraphics[width=4.8cm,height=4.8cm]{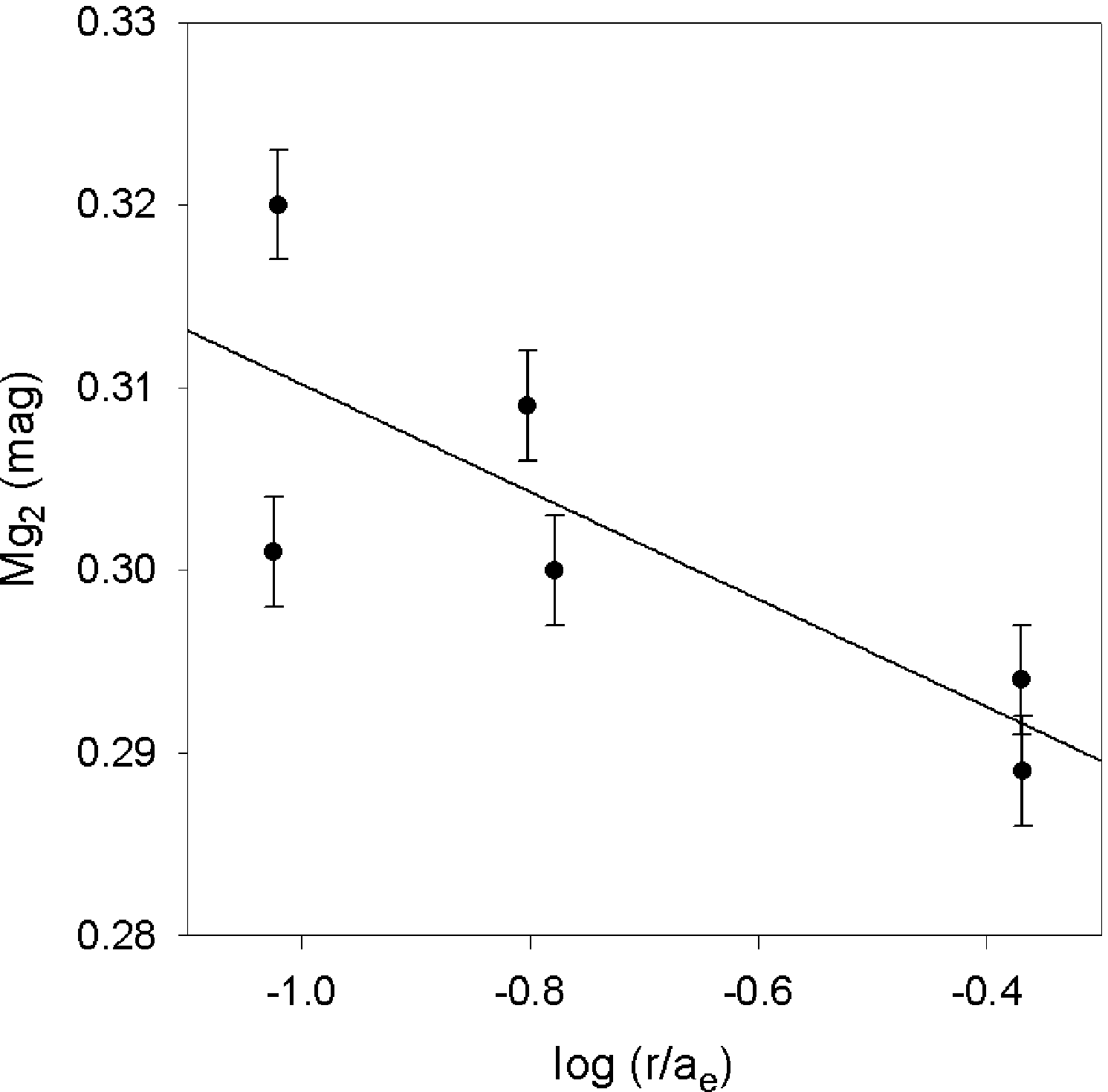}}\quad
         \subfigure[ESO488-027]{\includegraphics[width=4.8cm,height=4.8cm]{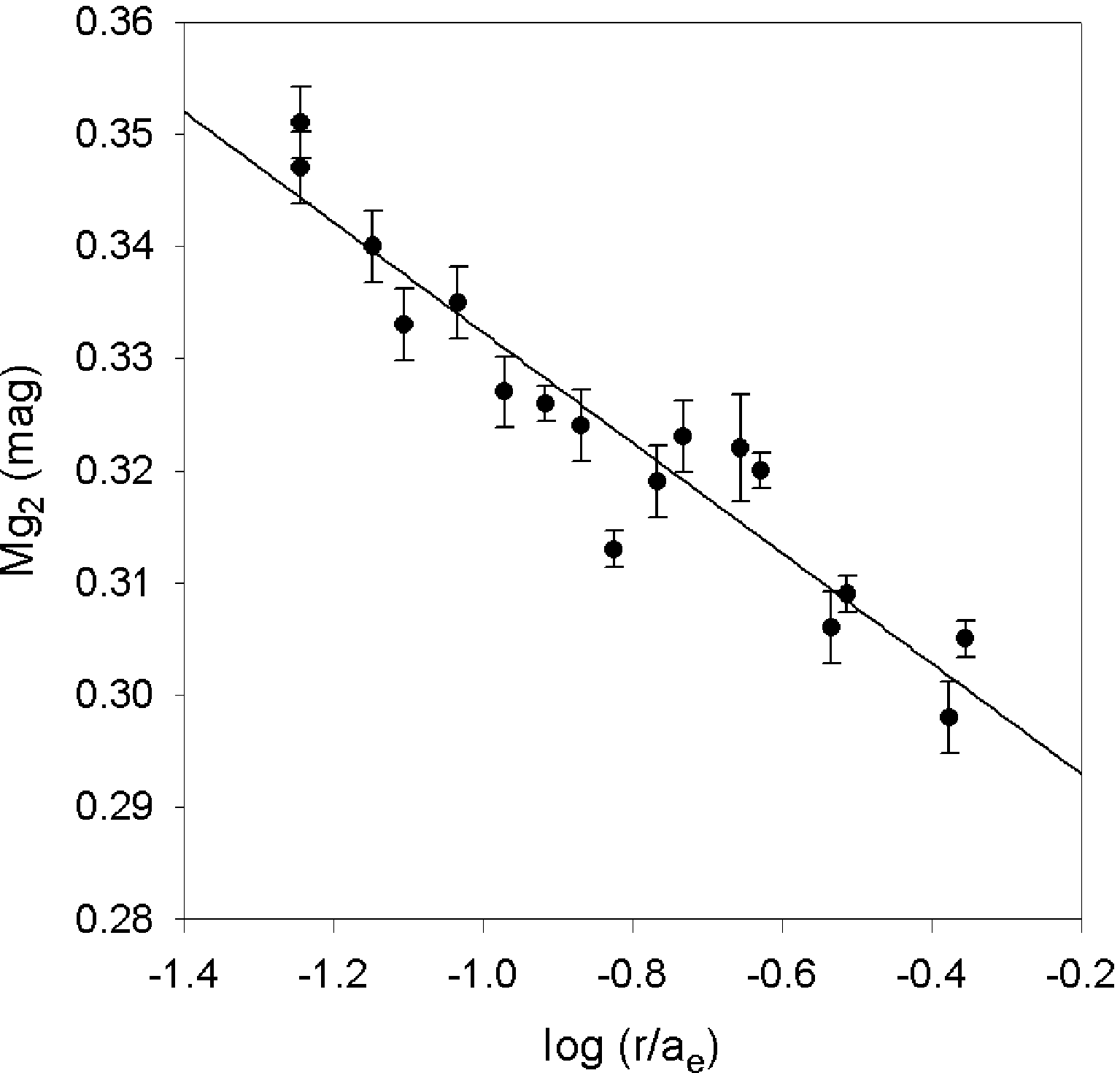}}}
 \mbox{\subfigure[ESO552-020]{\includegraphics[width=4.8cm,height=4.8cm]{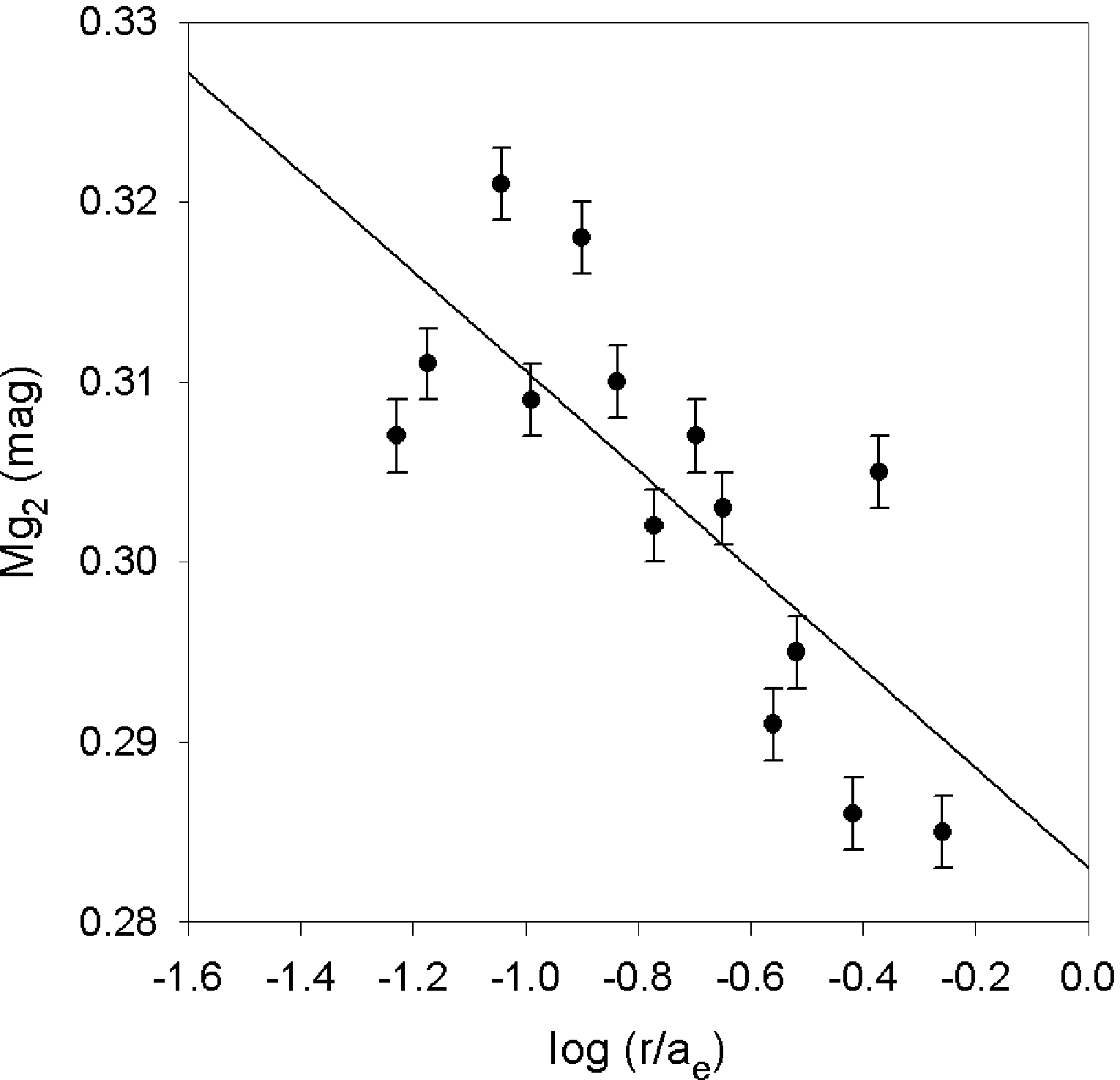}}\quad
         \subfigure[IC1633]{\includegraphics[width=4.8cm,height=4.8cm]{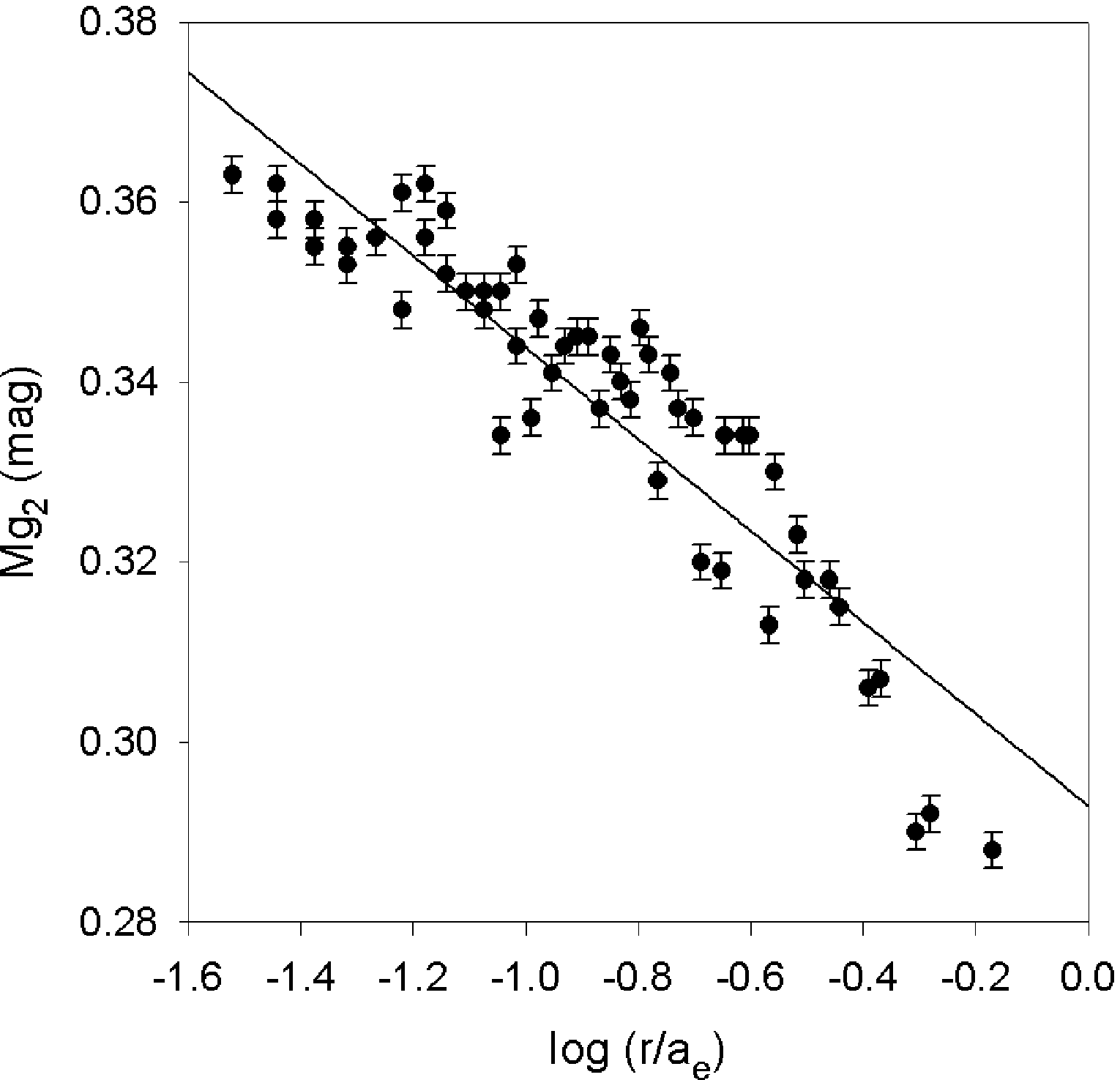}}\quad
         \subfigure[IC5358]{\includegraphics[width=4.8cm,height=4.8cm]{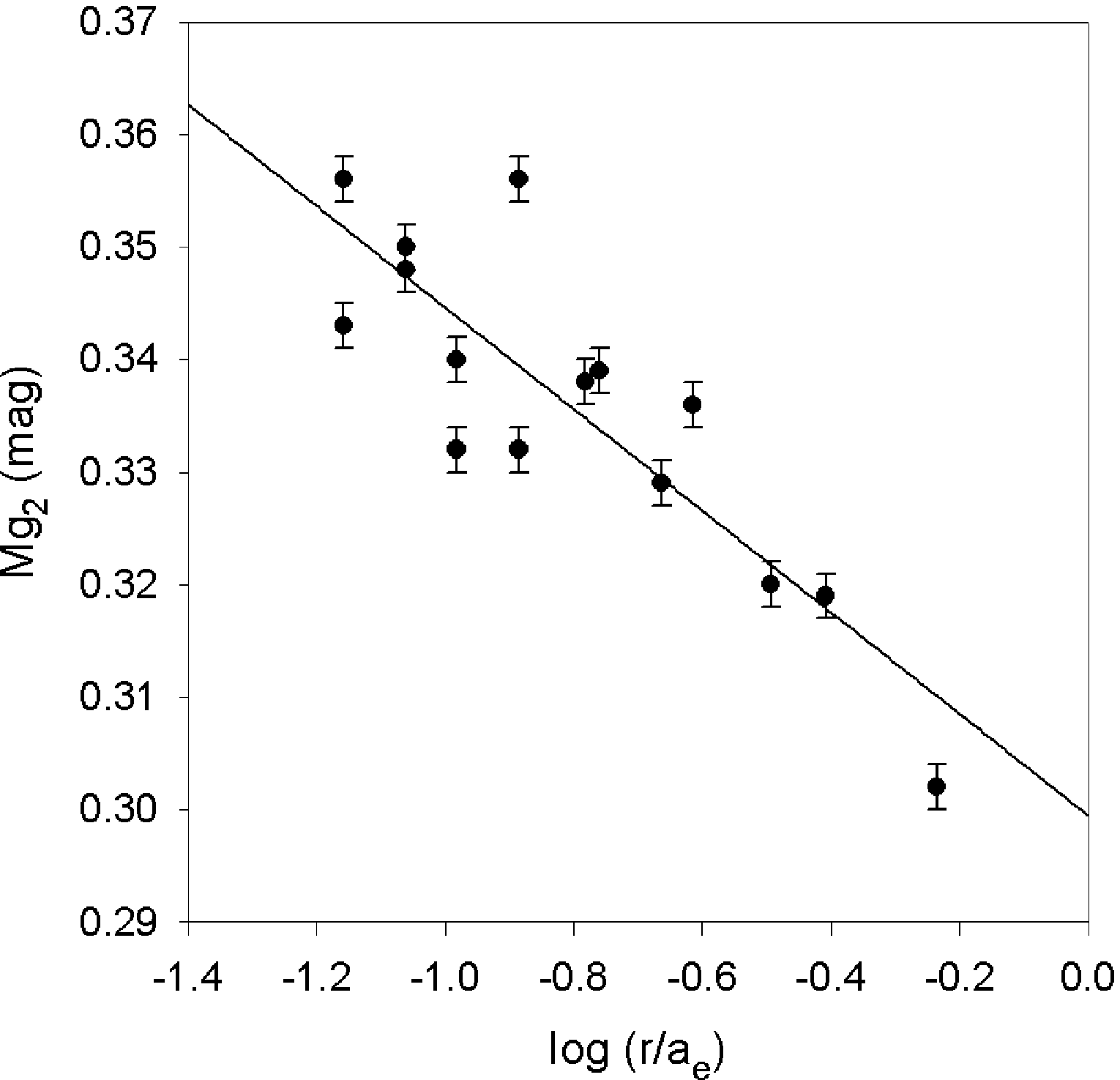}}}
 \mbox{\subfigure[MCG-02-12-039]{\includegraphics[width=4.8cm,height=4.8cm]{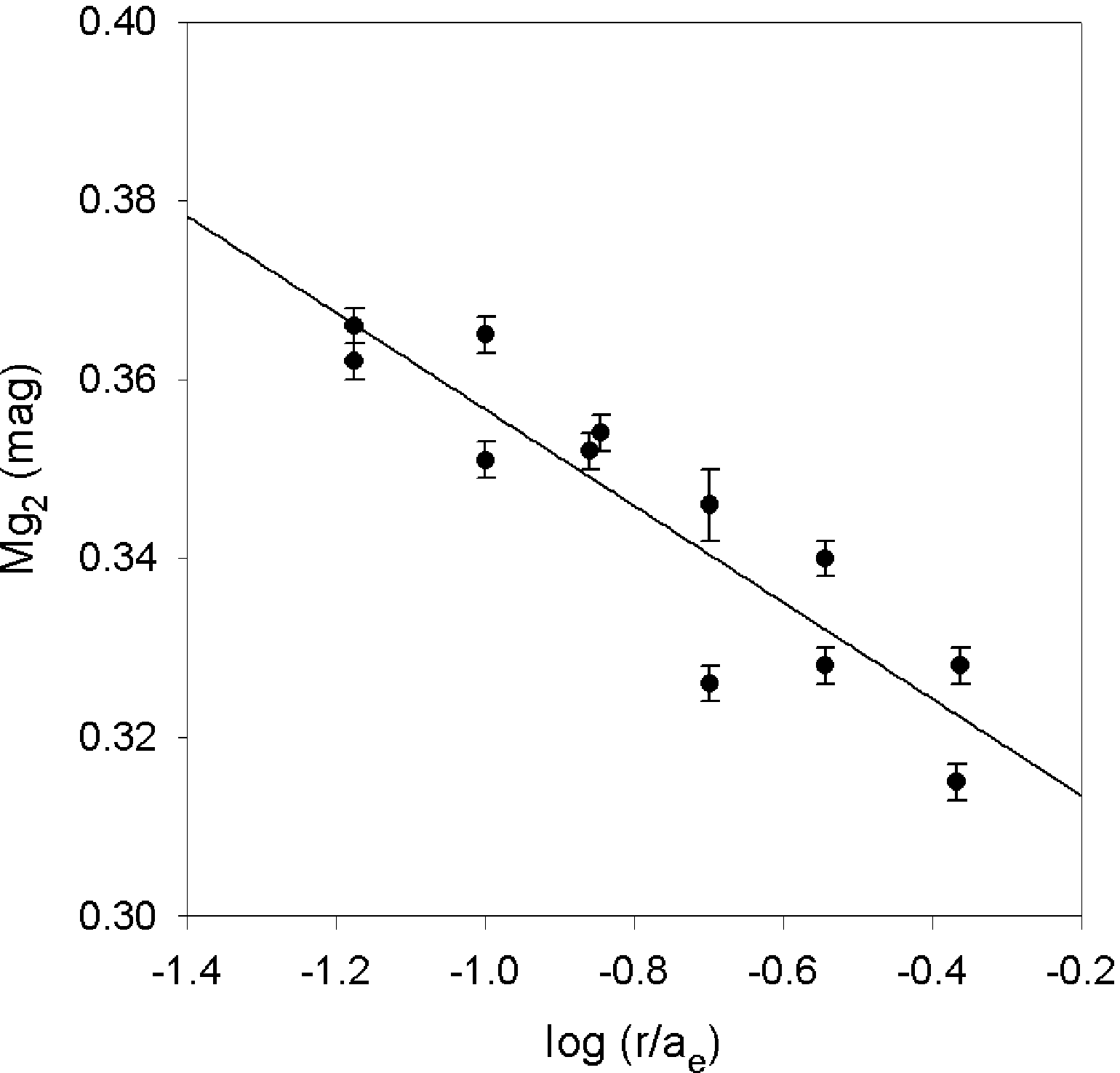}}\quad
         \subfigure[NGC0533]{\includegraphics[width=4.8cm,height=4.8cm]{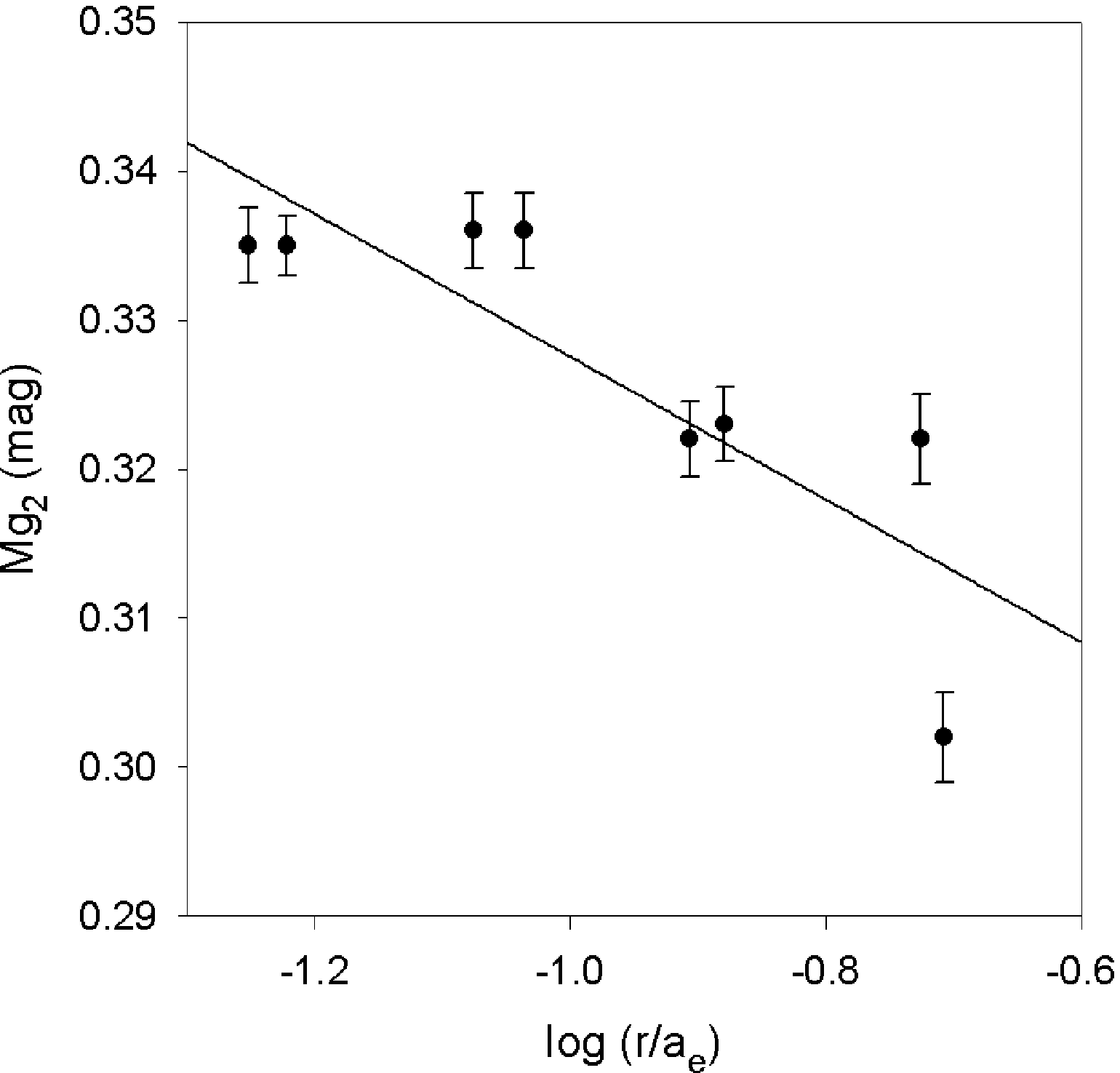}}\quad
         \subfigure[NGC0541]{\includegraphics[width=4.8cm,height=4.8cm]{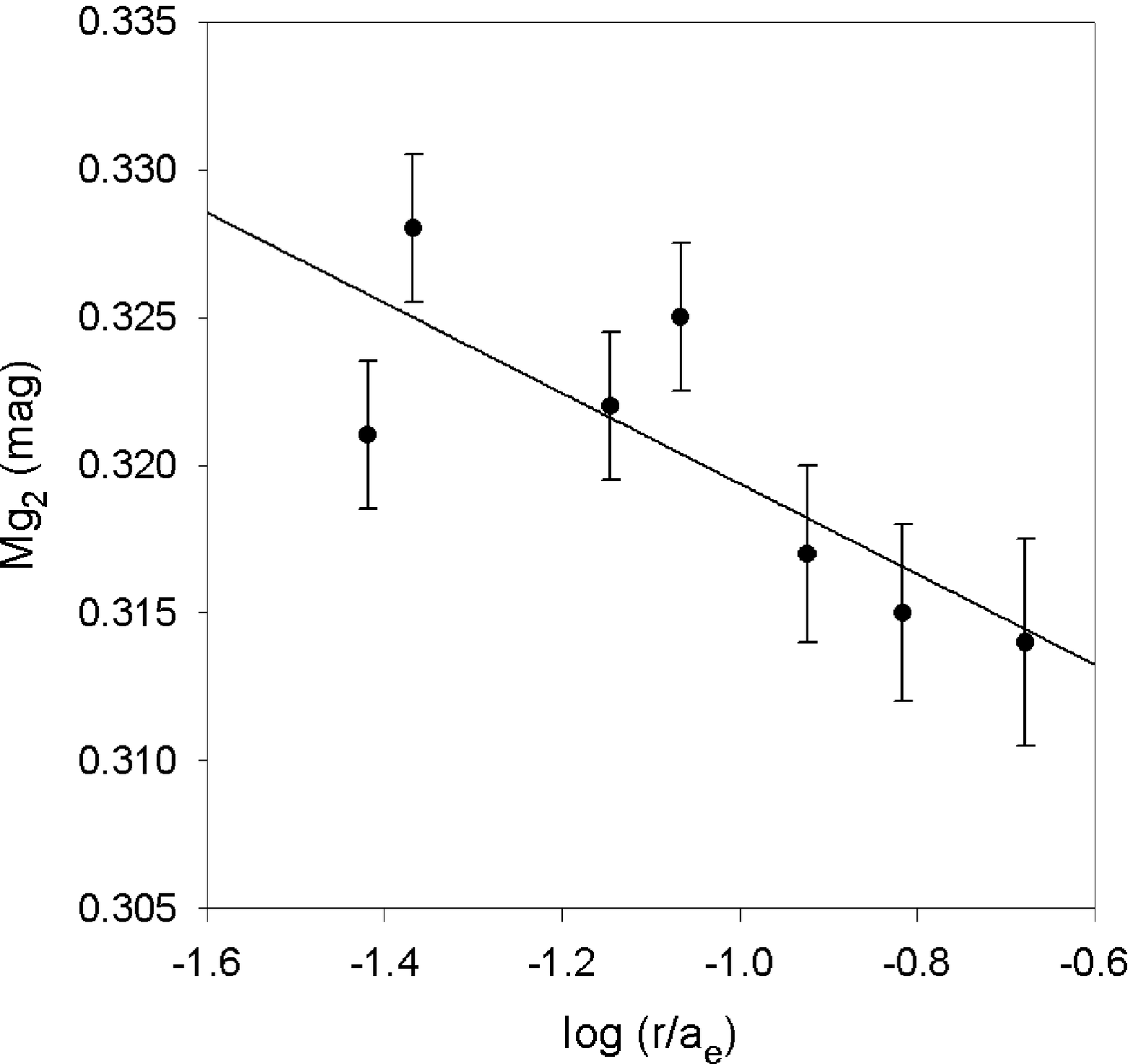}}}
\caption[]{BCG Mg$_{2}$ gradients. All data is folded with respect to the galaxy centres. The central 0.5 arcsec, to each side (thus in total the central 1 arcsec, comparable to the  seeing), were excluded in the figure as well as in the fitted correlations.}
   \label{Fig:Mg2_1}
\end{figure*}

\begin{figure*}
   \centering
\mbox{\subfigure[NGC1399]{\includegraphics[width=4.8cm,height=4.8cm]{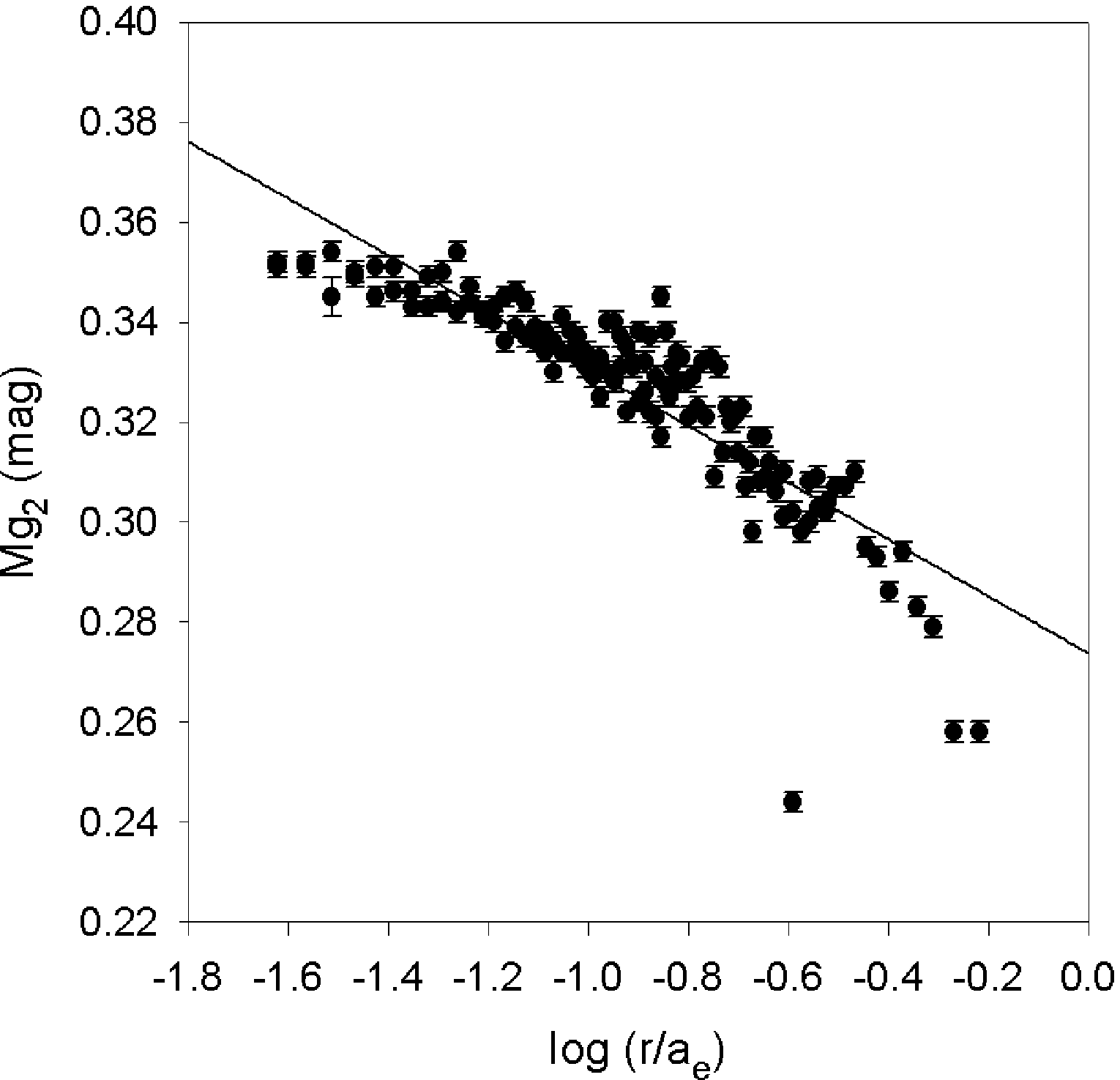}}\quad
         \subfigure[NGC1713]{\includegraphics[width=4.8cm,height=4.8cm]{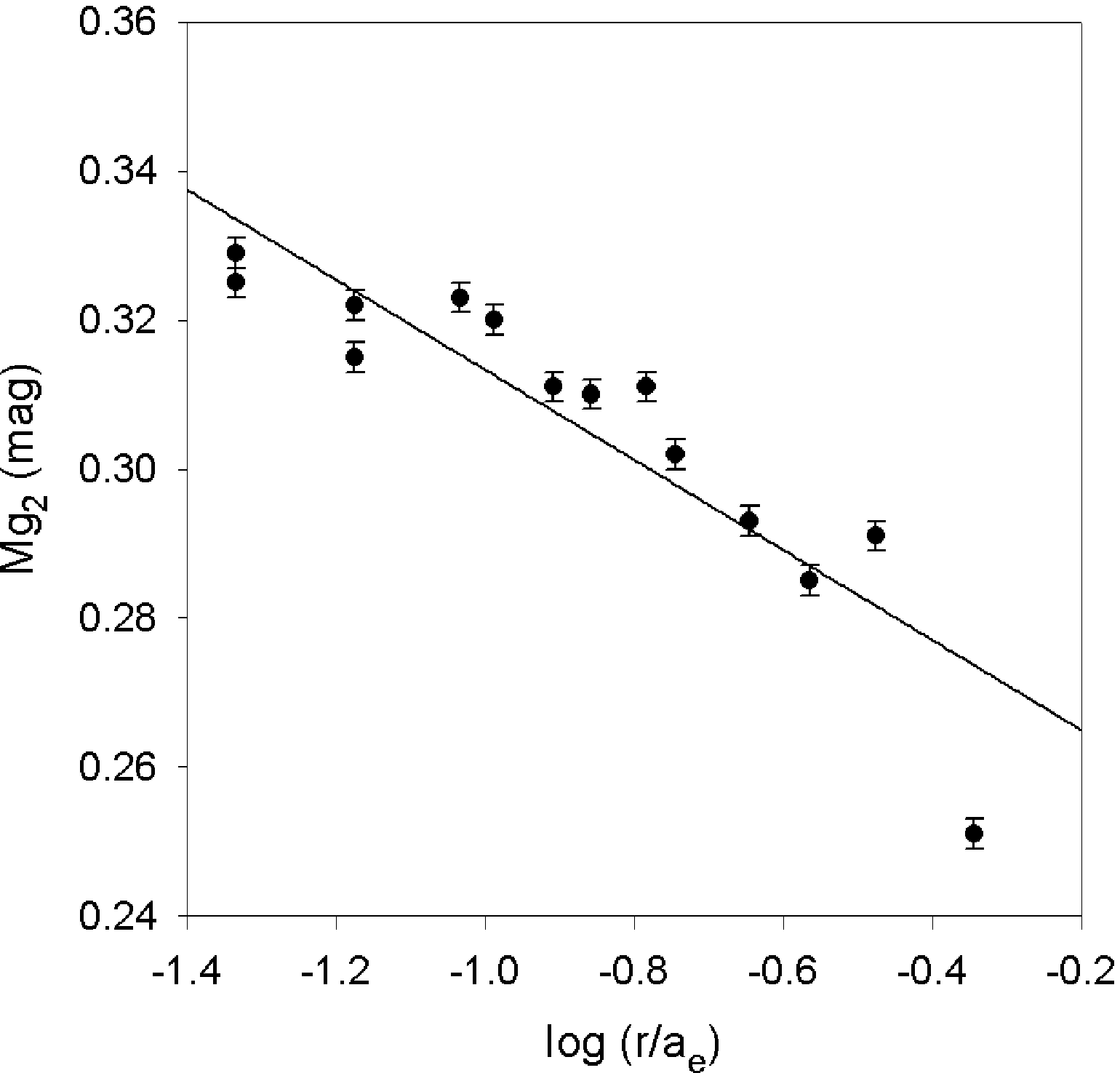}}\quad
         \subfigure[NGC2832]{\includegraphics[width=4.8cm,height=4.8cm]{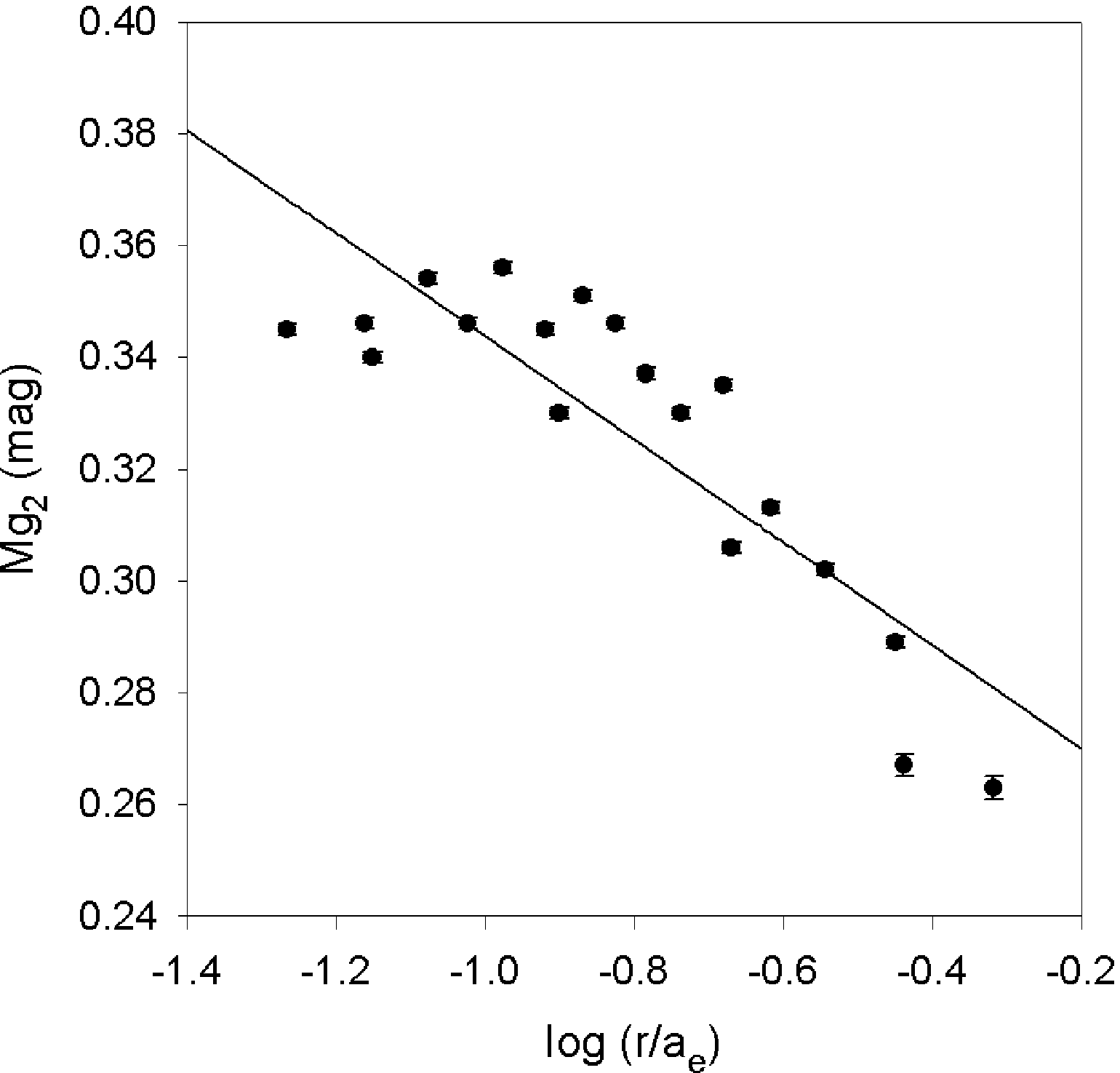}}}
\mbox{\subfigure[NGC4839]{\includegraphics[width=4.8cm,height=4.8cm]{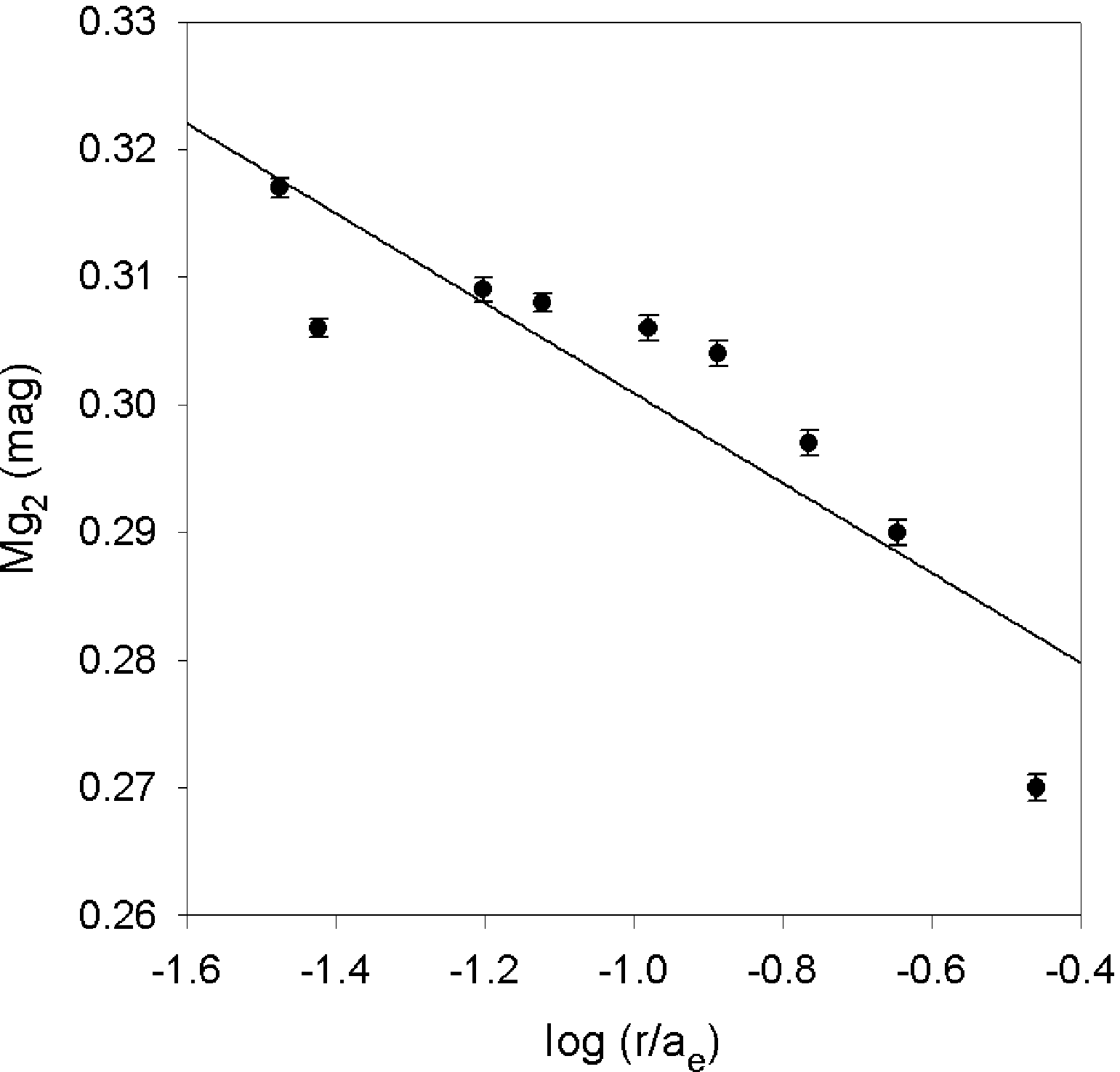}}\quad
         \subfigure[NGC6173]{\includegraphics[width=4.8cm,height=4.8cm]{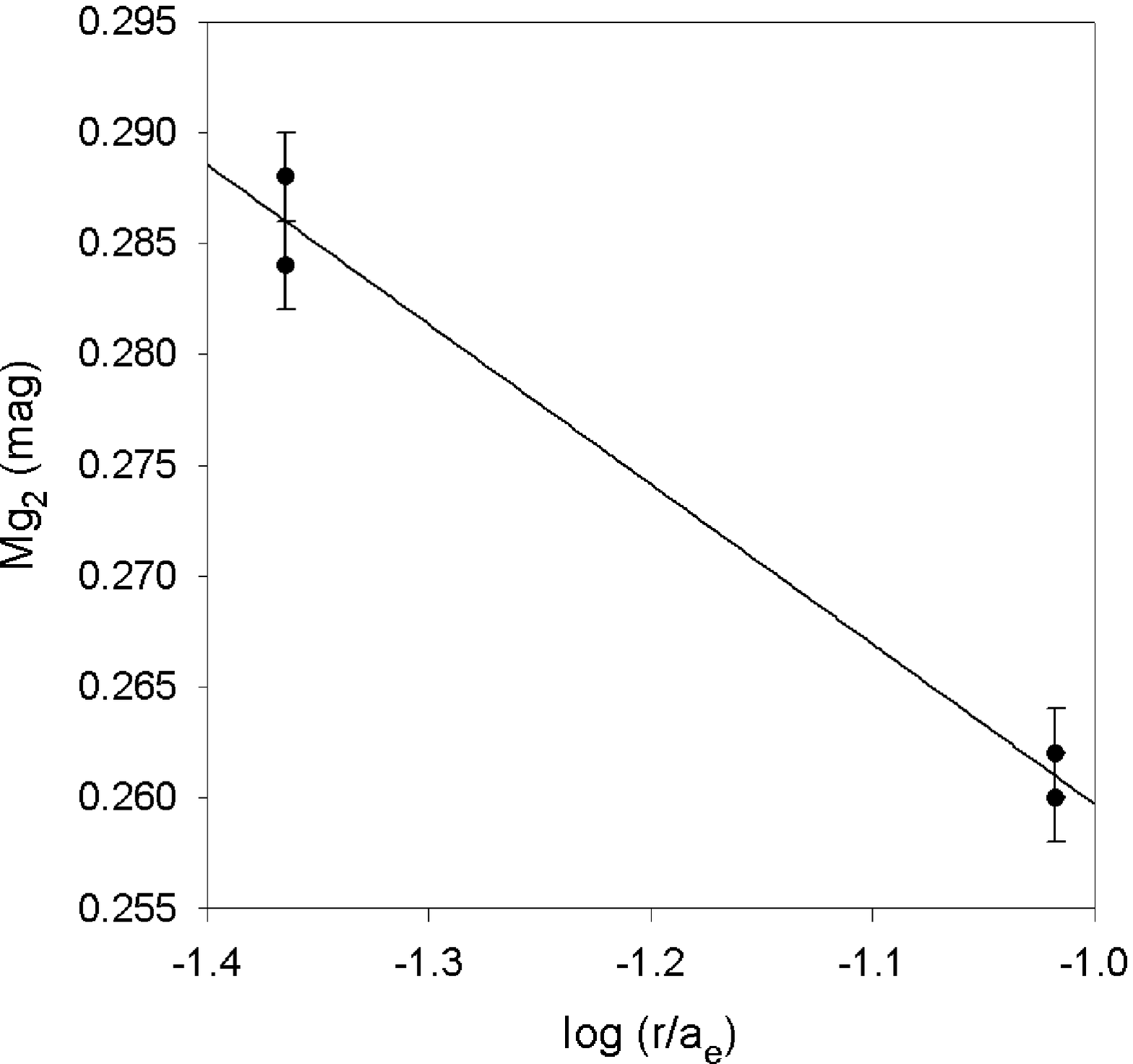}}\quad
         \subfigure[NGC6269]{\includegraphics[width=4.8cm,height=4.8cm]{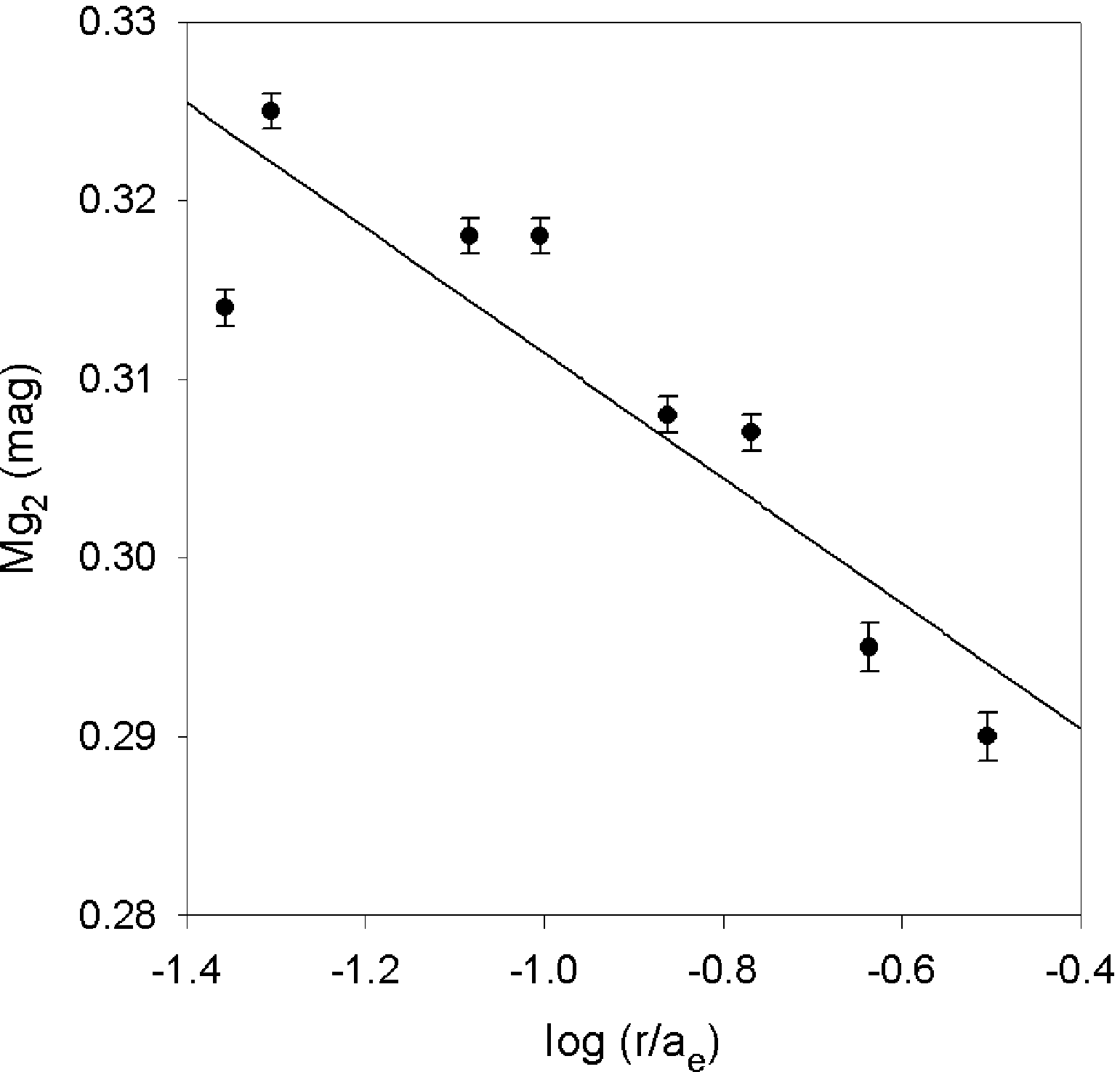}}}
\mbox{\subfigure[NGC7012]{\includegraphics[width=4.8cm,height=4.8cm]{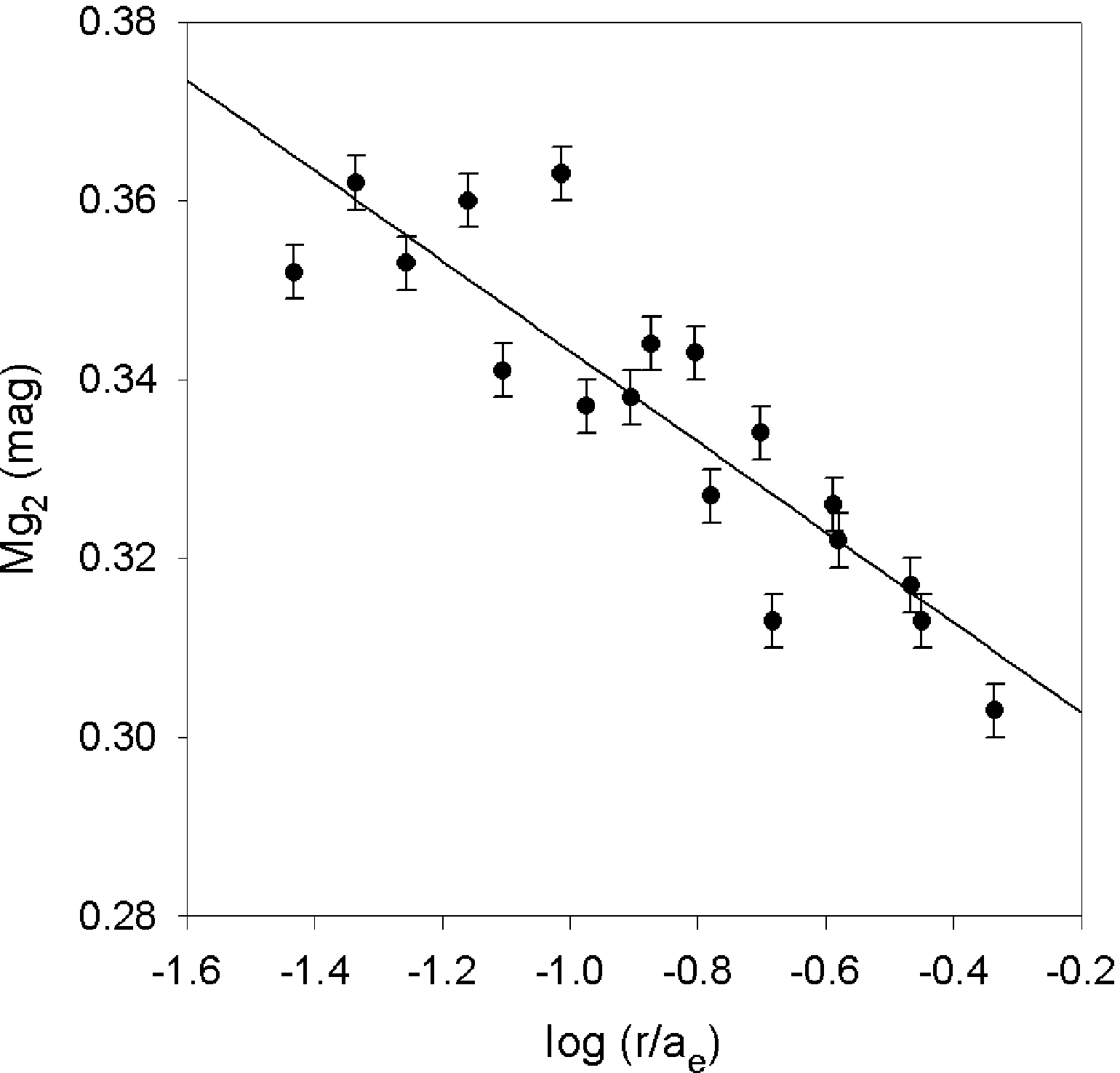}}\quad
         \subfigure[PGC004072]{\includegraphics[width=4.8cm,height=4.8cm]{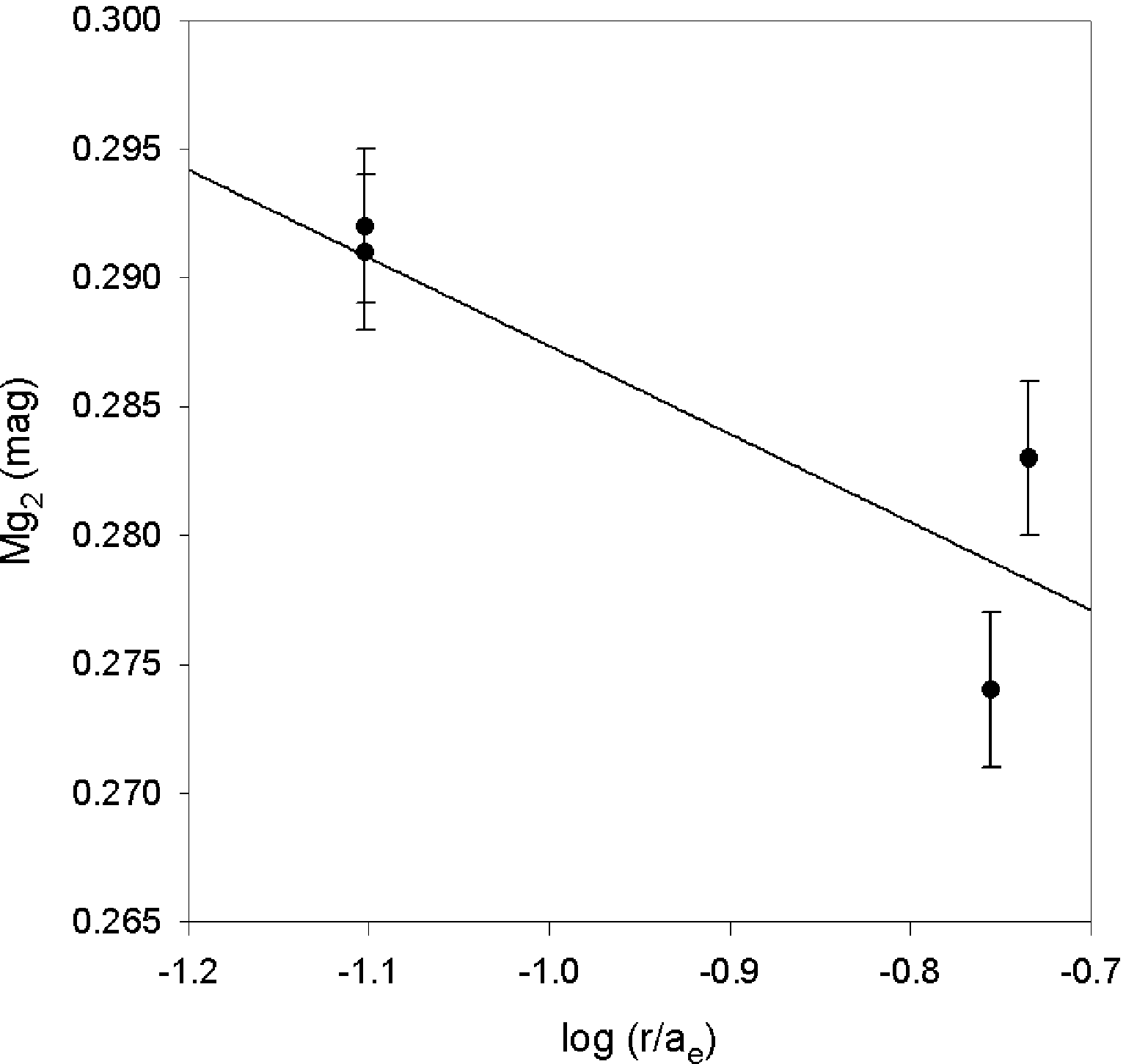}}\quad
         \subfigure[PGC030223]{\includegraphics[width=4.8cm,height=4.8cm]{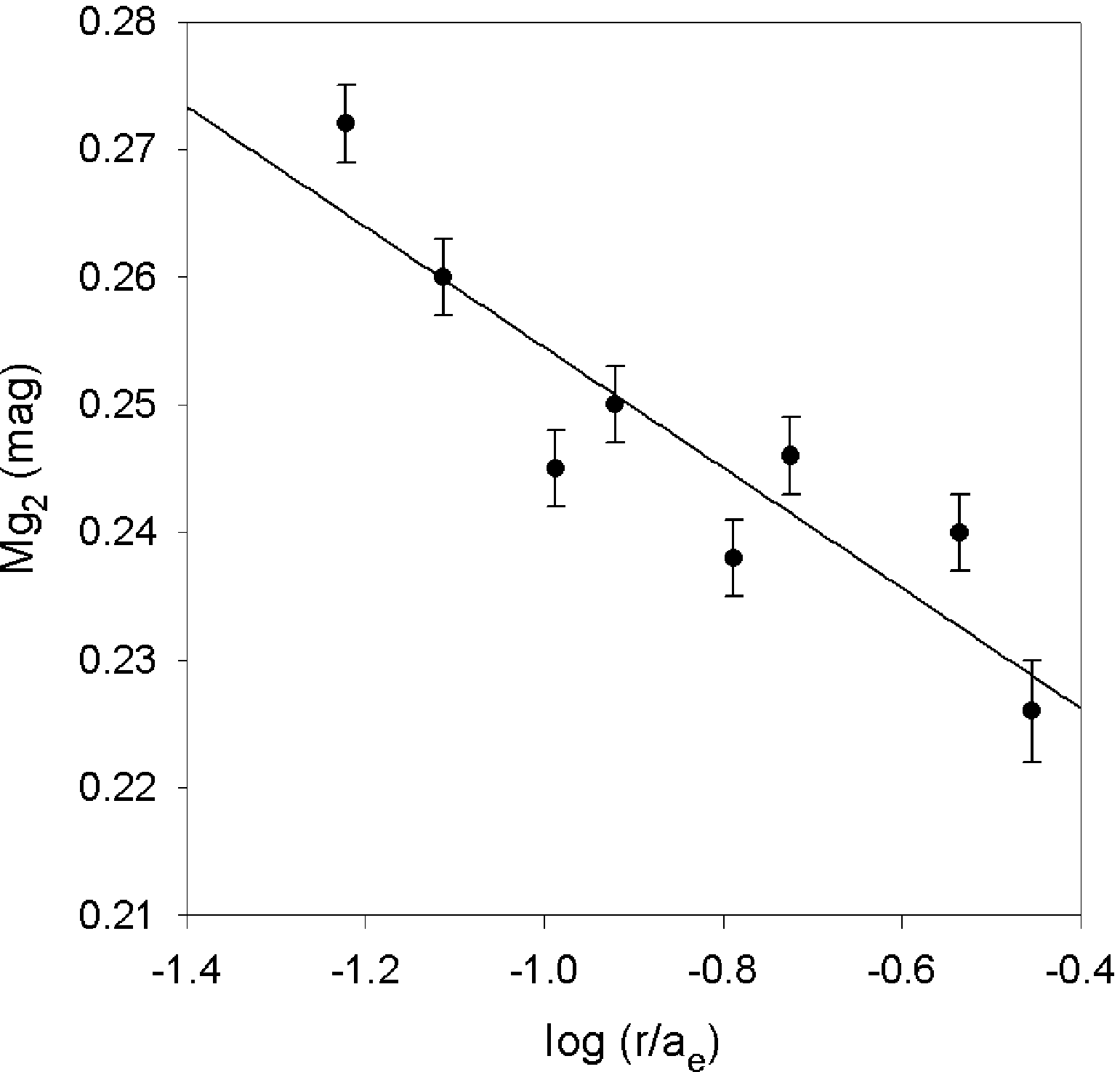}}}
\mbox{\subfigure[PGC072804]{\includegraphics[width=4.8cm,height=4.8cm]{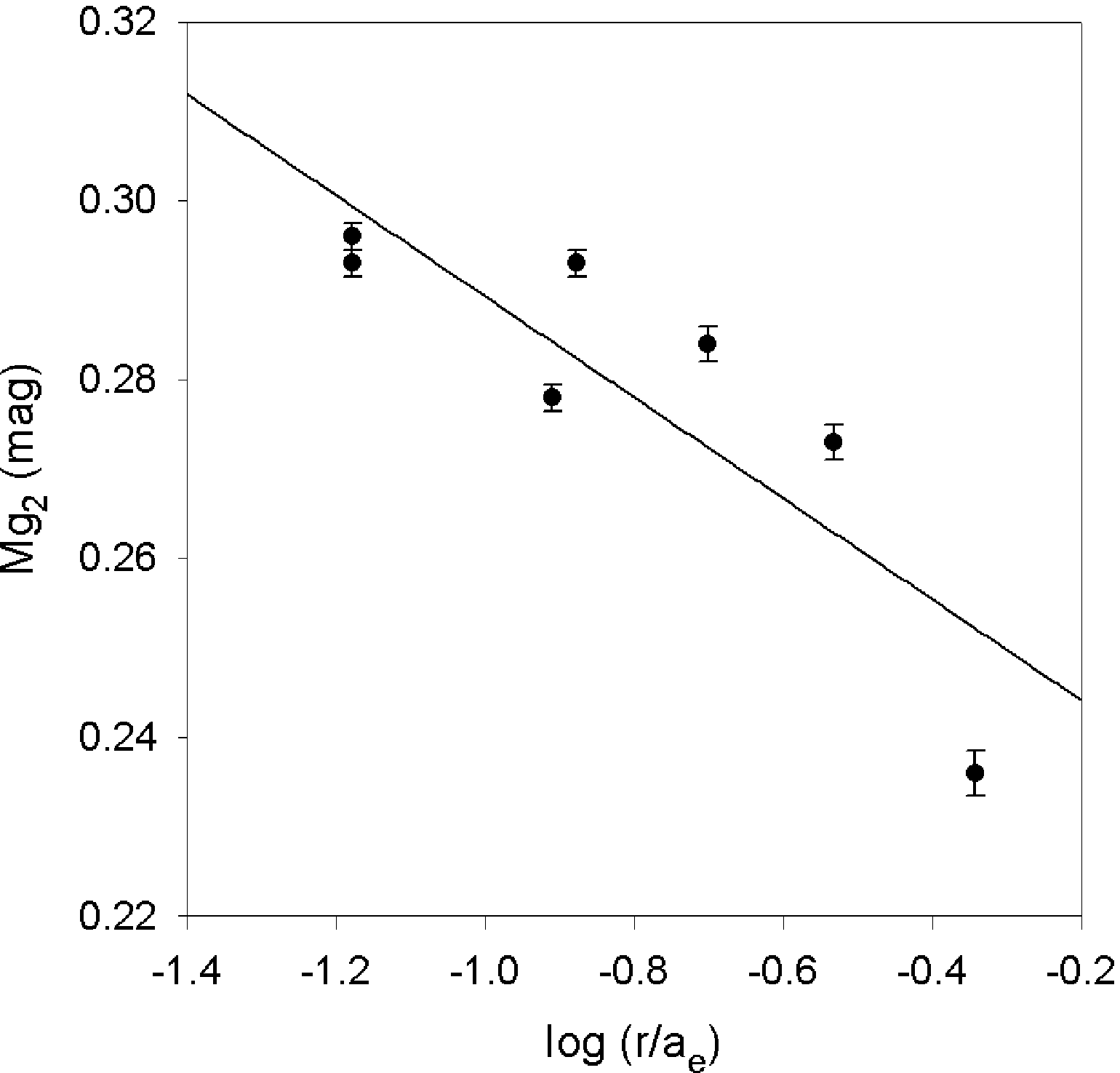}}\quad
         \subfigure[UGC02232]{\includegraphics[width=4.8cm,height=4.8cm]{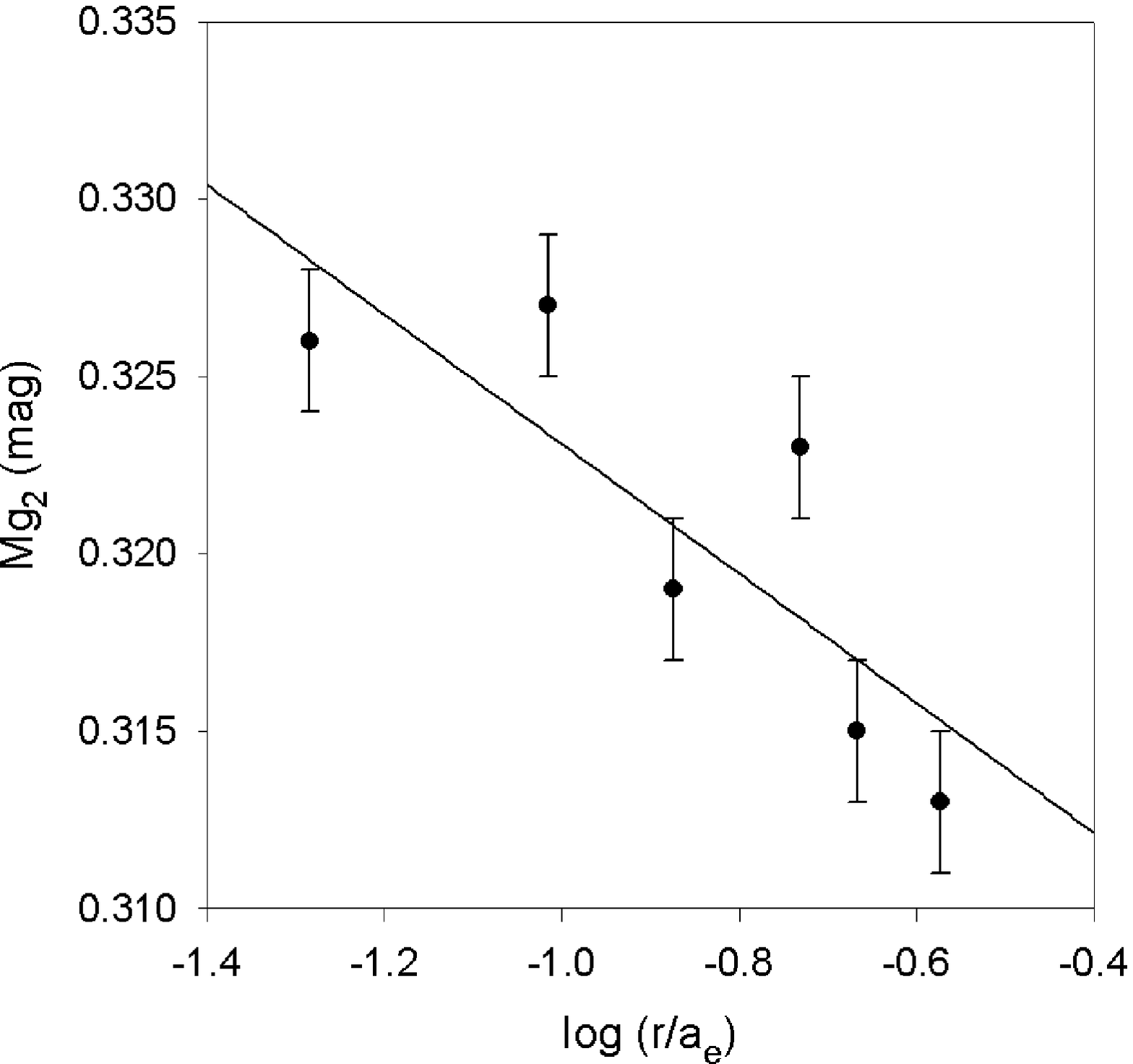}}\quad
\subfigure[UGC05515]{\includegraphics[width=4.8cm,height=4.8cm]{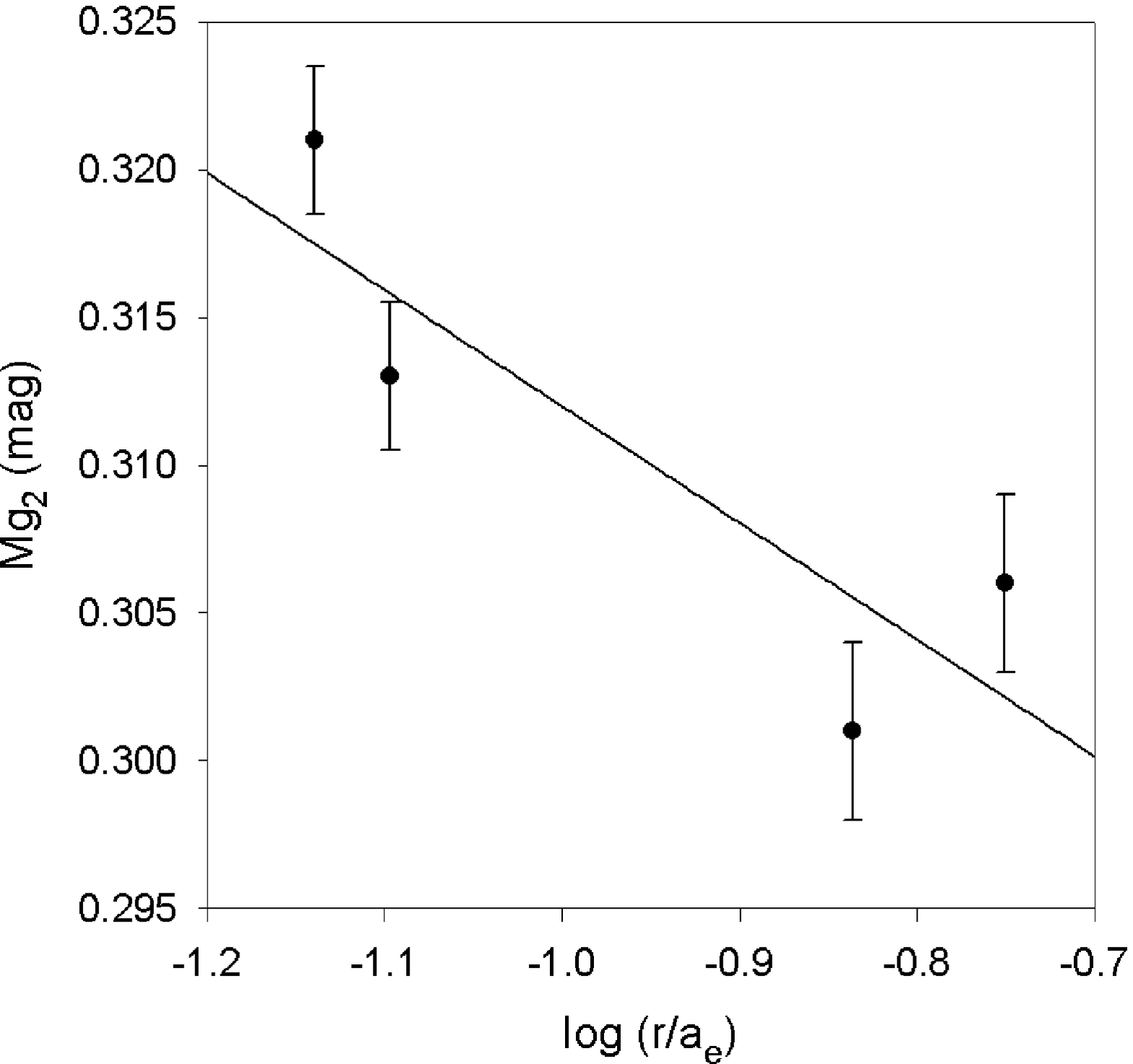}}}
\caption[]{BCG Mg$_{2}$ gradients continue.}
   \label{Fig:Mg2}
\end{figure*}

\begin{figure*}
   \centering
\mbox{\subfigure[GSC555700266]{\includegraphics[width=4.1cm,height=4.1cm]{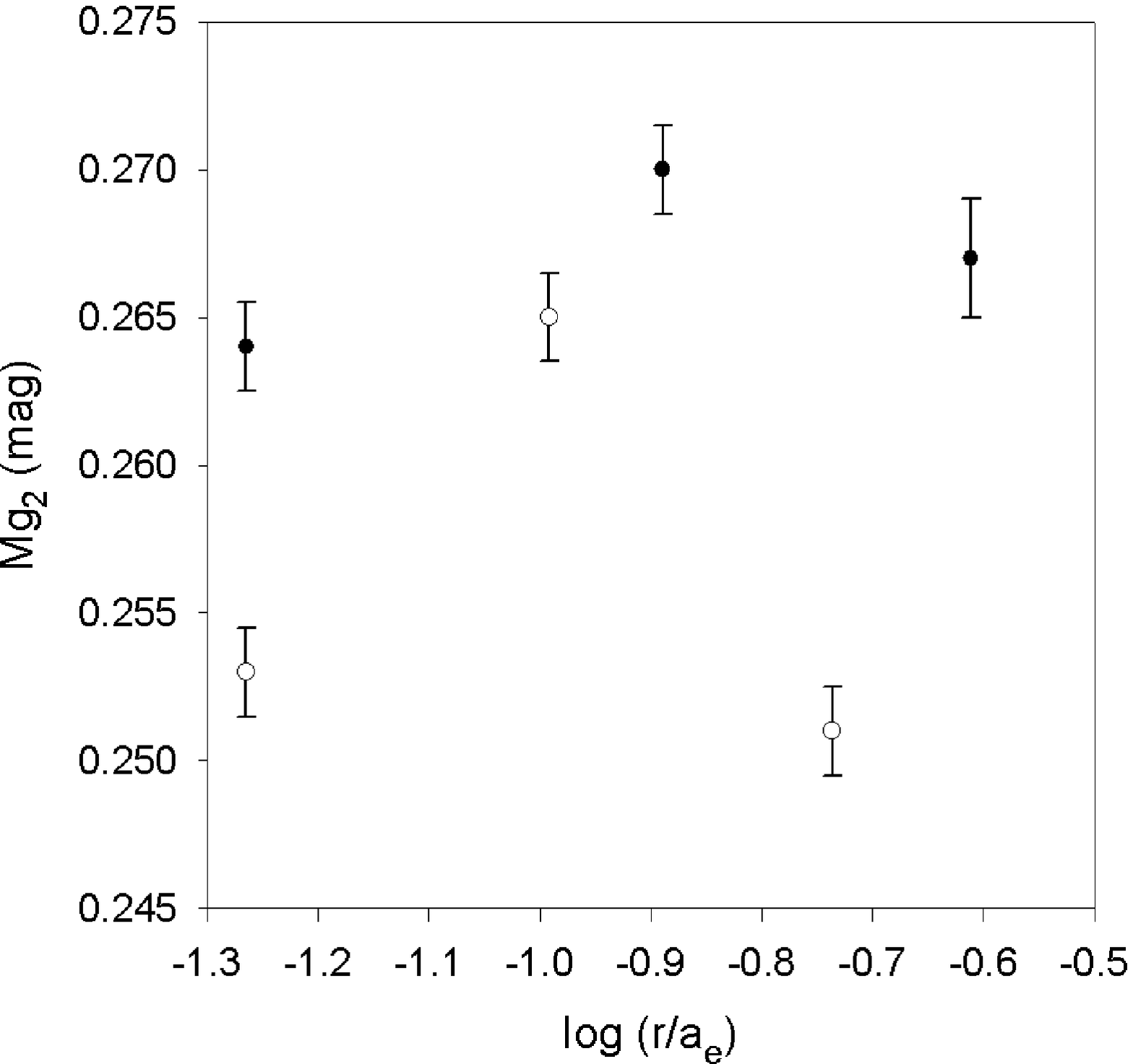}}\quad
         \subfigure[IC4765]{\includegraphics[width=4.1cm,height=4.1cm]{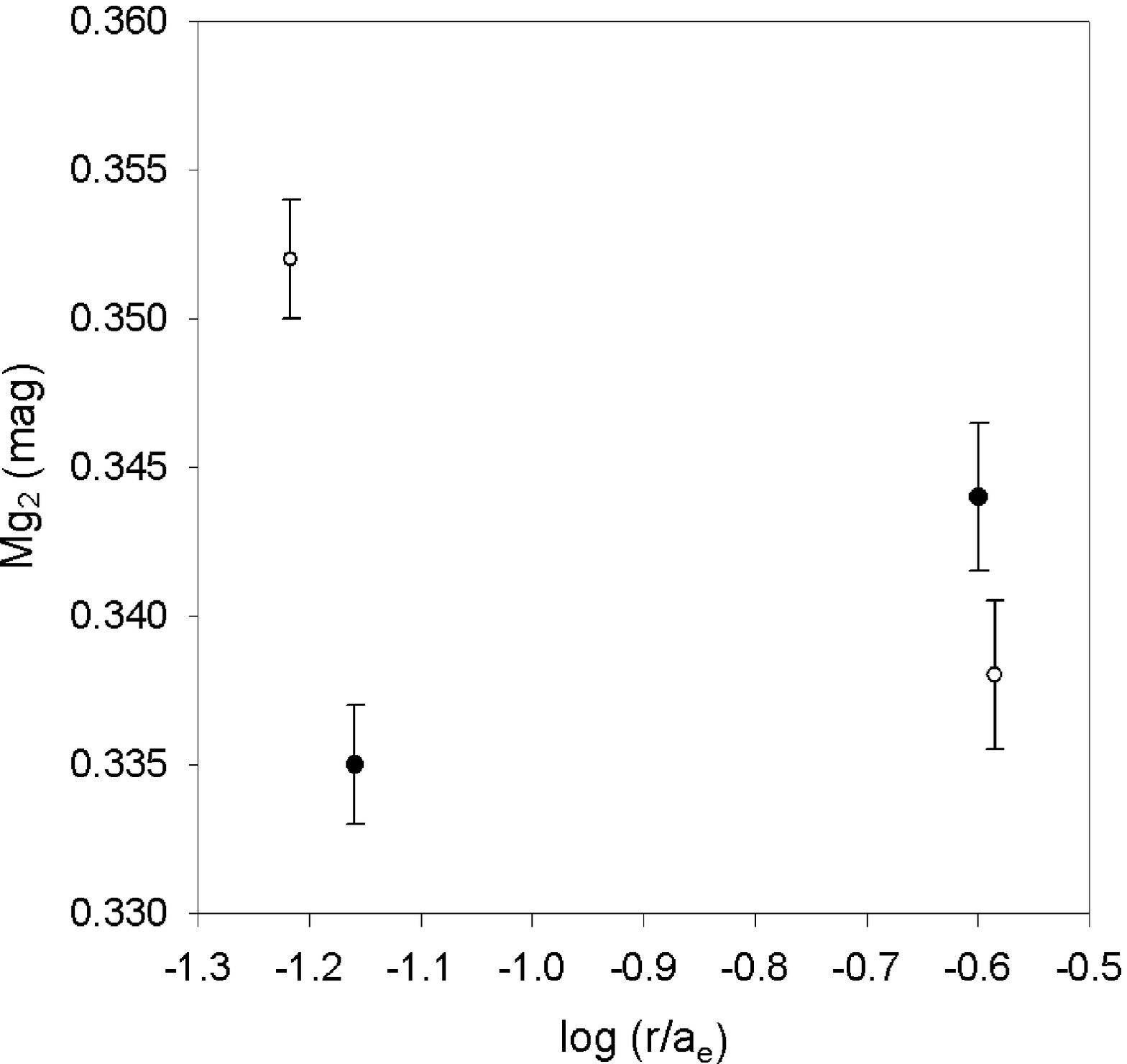}}\quad
         \subfigure[Leda094683]{\includegraphics[width=4.1cm,height=4.1cm]{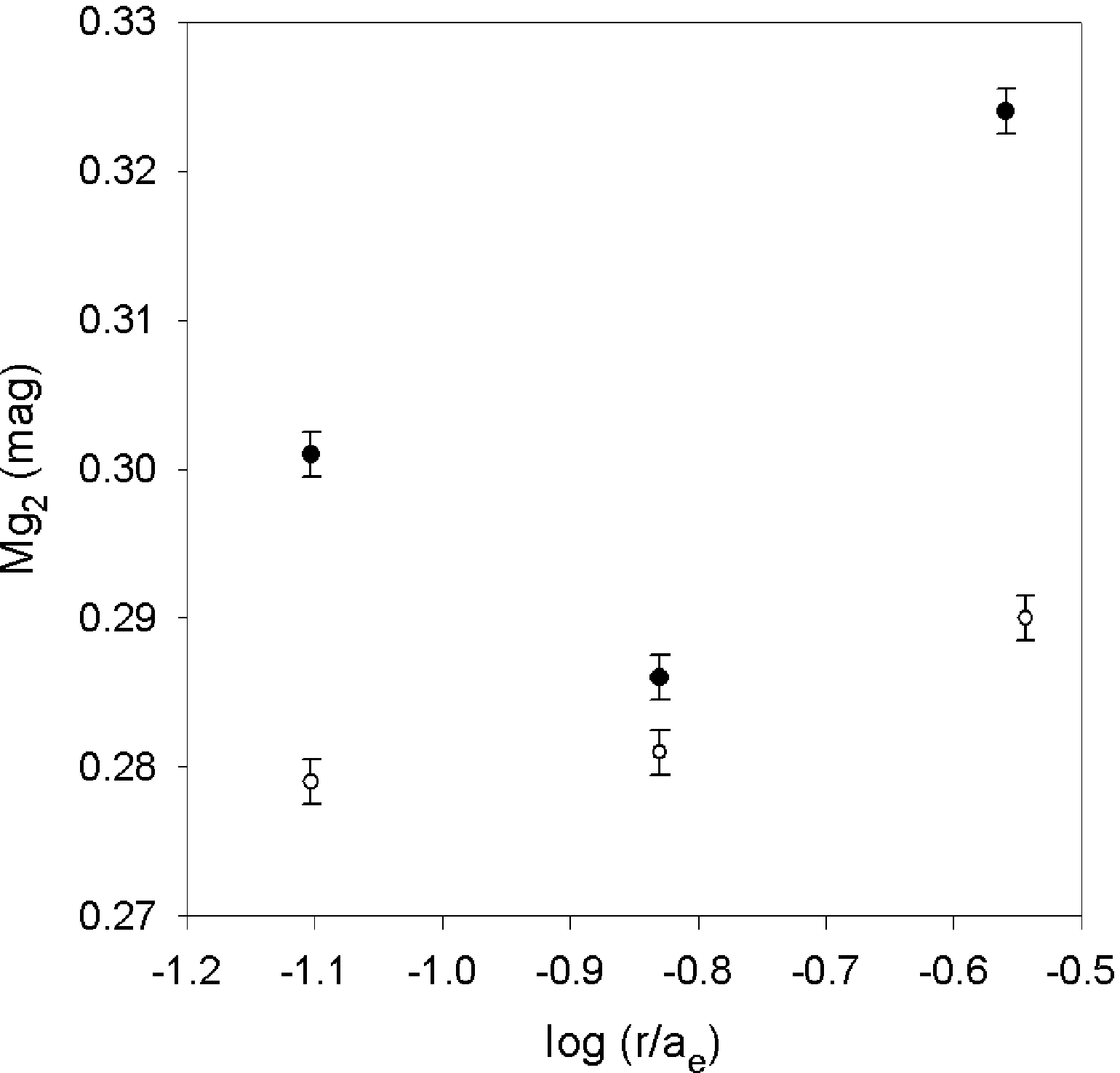}}\quad
         \subfigure[NGC3311]{\includegraphics[width=4.1cm,height=4.1cm]{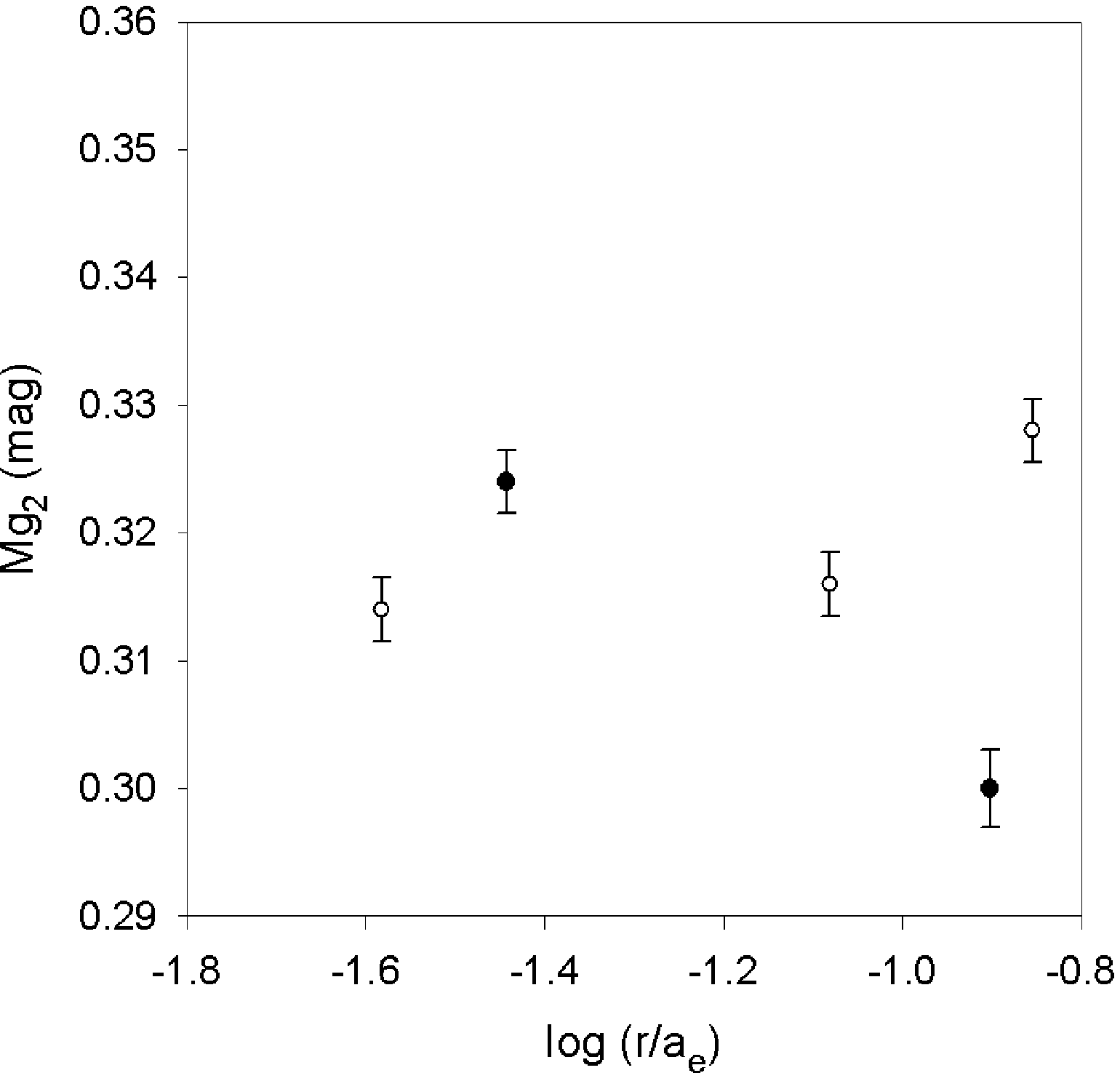}}}
\caption[]{The asymmetric BCG Mg$_{2}$ gradients. The black and empty symbols indicate the two different sides of the galaxy.}
   \label{Fig:assym}
\end{figure*}

We primarily investigate the gradients in the Mg$_{2}$ index here since it is sensitive to metallicity, and has been exhaustively studied for elliptical galaxies by previous authors. It eliminates the systematics (such as different stellar libraries and minimisation techniques) introduced by fitting stellar population models. It is also known that stellar population parameters, in particular metallicity, depends strongly on the indices used to determine it (S\'{a}nchez-Bl\'{a}zquez et al.\ 2006b, Pipino et al.\ 2010). We will transform the index measurements into age and metallicity gradients in Loubser et al.\ (in preparation).

We take the gradient as the slope of a linear fit to the relation $I^{'}=a+b\log(\frac{r}{a_{e}})$ where $I^{'}$ is the Mg$_{2}$ index in magnitudes and $a_{e}$ is the effective radius as presented in Table \ref{table:objects}. Gradients are taken to be significantly different from zero if the slope is greater than three times the 1$\sigma$ error on that gradient. Four galaxies in our sample have gradients that are compatible with being null, within the errors. Note that, despite our derived gradients reaching on average a fraction of 0.4 of the $a_{e}$ of the galaxy, Brough et al.\ (2007) showed that Mg$_{2}$ gradients in BCGs do not deviate from a power law up to three times $a_{e}$. Therefore, our results are valid even if very large radii are not reached. 

We plot the Mg$_{2}$ gradients in Figures \ref{Fig:Mg2_1} and \ref{Fig:Mg2}. The Mg$_{2}$ gradients of GSC555700266, IC4765, Leda094683, and NGC3311 were not fitted since the two sides of the galaxies showed asymmetrical profiles (the asymmetric profiles are shown in Figure \ref{Fig:assym}, where the two galaxy sides are indicated by the black and empty symbols). We have investigated these four galaxies (with respect to their emission, slit placement, radial kinematic profiles (see paper 2), central stellar populations and host cluster X-ray properties) without finding any reason for the asymmetric Mg$_{2}$ gradients. Since these galaxies have few data points, the two galaxy sides could not be fitted separately, and they were excluded from further analyses.

\begin{figure}
   \centering
   \includegraphics[scale=0.7]{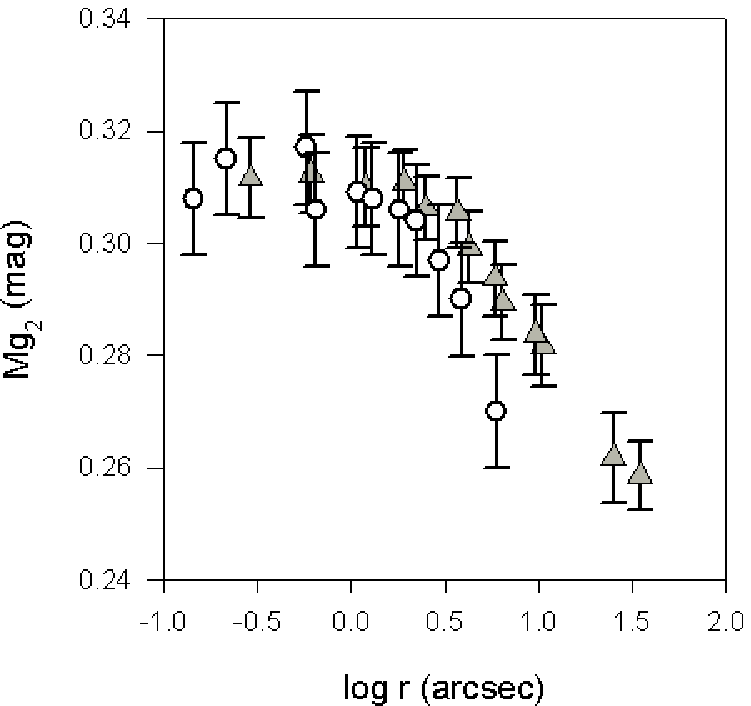}
   \caption{NGC4839: Comparison to previous data from Mehlert et al.\ (2000 -- grey triangles). The data are folded with respect to the centre of the galaxy.}
   \label{fig:Previous_comp}
\end{figure}

The Mg$_{2}$ gradients of one of the galaxies could be compared to previous measurements in the literature from Mehlert et al.\ (2000). The comparison is shown in Figure \ref{fig:Previous_comp}. The profiles compare well and small differences can be as a result of spectral resolution and slit width.

We find a mean gradient ($dI^{'}/\log(\frac{r}{a_{e}})$) of --0.047$\pm$0.009 mag for the 21 BCG Mg$_{2}$ gradients, and --0.051$\pm$0.004 mag, for the 17 BCGs with significant (non-zero) Mg$_{2}$ gradients. This is consistent with previous studies of both BCGs, and massive ellipticals (--0.053 $\pm$ 0.015 by Couture $\&$ Hardy 1988 for six ellipticals; --0.058 $\pm$ 0.020 by Gorgas, Efstathiou $\&$ Arag\'{o}n-Salamanca 1990 for 16 ellipticals; --0.059 $\pm$ 0.022 by Davies, Sadler $\&$ Peletier 1993 for 13 ellipticals; --0.058 by Cardiel et al.\ 1998a for eight BCGs). For NGC6173 our Mg$_{2}$ gradient of --0.072$\pm$0.006 also agrees well with that measured by Carter et al.\ (1999), who found --0.073$\pm$0.002.    

\subsection{Mg$_{2}$ gradient -- velocity dispersion relation}

Up to now, there has been contradicting results among different studies about the existence of a relation between the Mg$_{2}$ gradients and velocity dispersion for elliptical galaxies (Peletier et al.\ 1990; Gorgas et al.\ 1990; Gonz\'{a}lez 1993; Davies et al.\ 1993; Carollo $\&$ Danziger 1994; Mehlert et al.\ 2003). However, over the last few years, authors have come to agree that this was due to the different behaviour that this relation has in different ranges of galaxy mass (see Carollo, Danziger $\&$ Buson 1993; S\'{a}nchez-Bl\'{a}zquez et al. 2007; Spolaor et al. 2009; and Kuntschner et al. 2010). For elliptical galaxies, gradients get steeper with mass for galaxies with masses below $\sim$10$^{11}$ M$_{\odot}$, while the opposite behaviour is observed for more massive galaxies (although the slope is much flatter and the scatter also increases). This behaviour was already noticed by Vader et al.\ (1988) using colour gradients. Physically, the transition point may mark the mass above which galaxy formation histories are dominated by late-epoch, gas-poor merging, which dilutes the metallicity gradient (S\'{a}nchez-Bl\'{a}zquez et al. 2007; Hopkins et al.\ 2009; Rawle, Smith $\&$ Lucey 2010). This transition also corresponds to the separation of rapidly and slowly rotating, and more or less to the boxy/disky, core/power-law separation, where the correlation inverts and gradients become flatter with mass.

\begin{figure}
   \centering
\mbox{\subfigure{\includegraphics[scale=0.65]{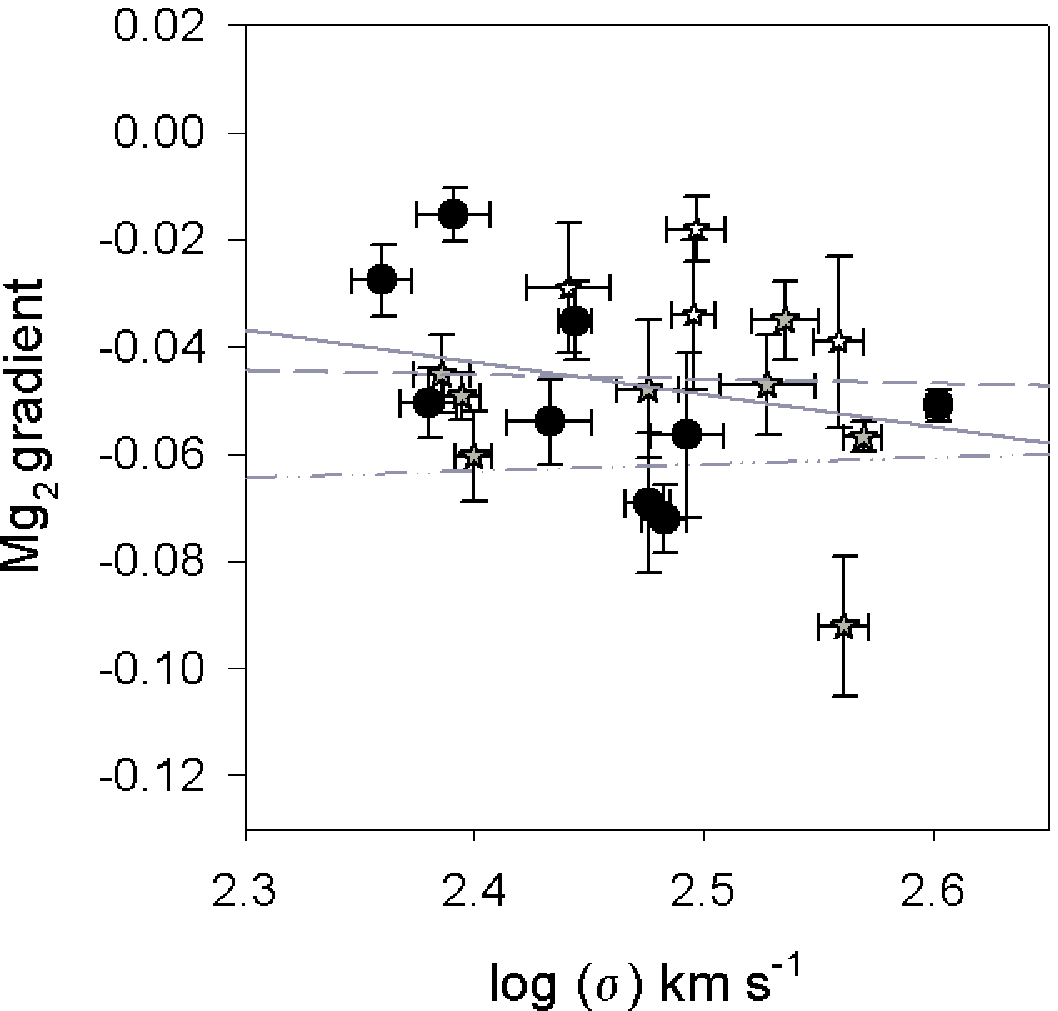}}}
         \mbox{\subfigure{\includegraphics[scale=0.65]{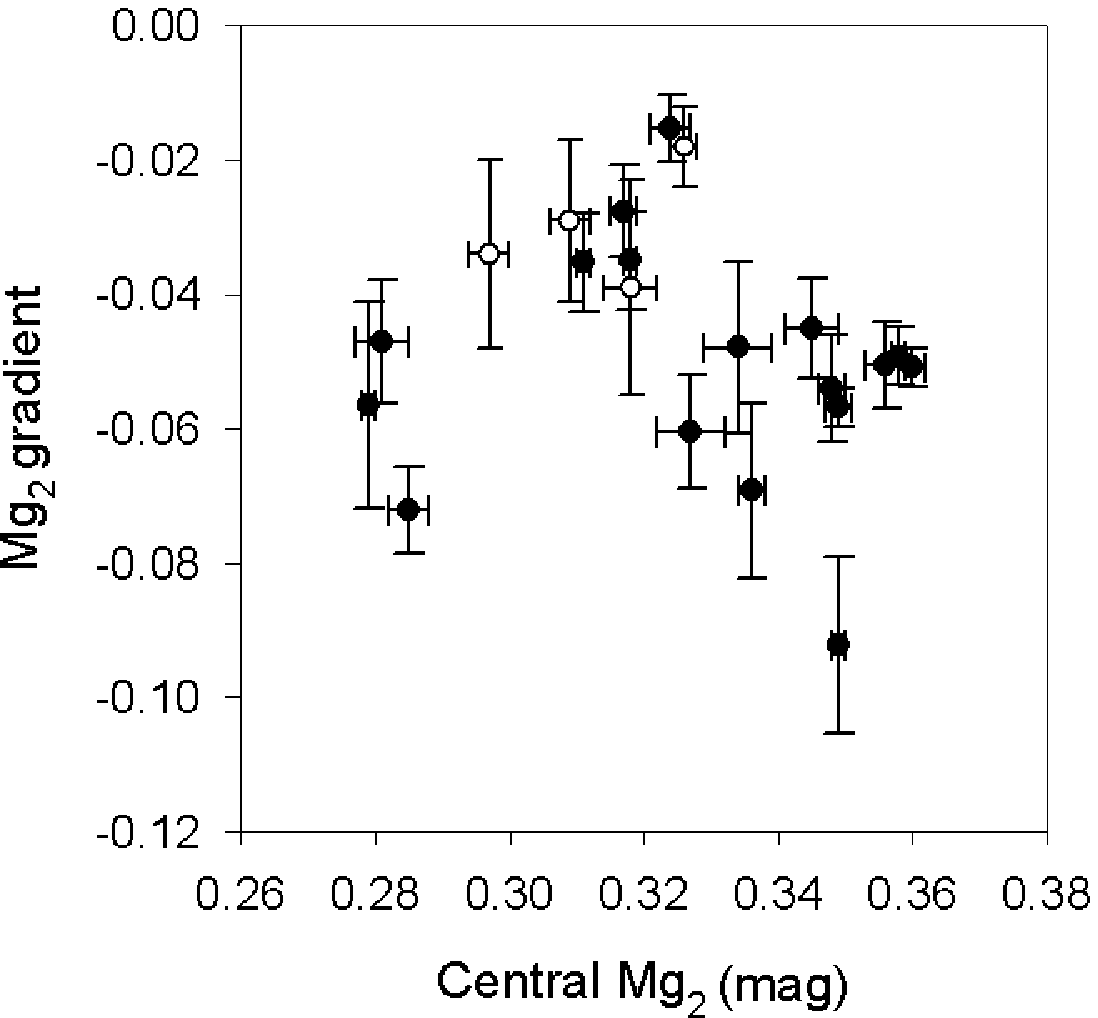}}}
   \caption{Mg$_{2}$ gradients plotted against central velocity dispersion and central Mg$_{2}$ measurements. The Mg$_{2}$ gradients where the PA was placed within 15 degrees of the MA (where the MA was known) are shown with filled circles, and the rest of gradients with grey stars (with the exception of the four non-significant gradients which are shown with empty stars). The solid grey line shows the Mg$_{2}$ gradient -- velocity dispersion correlation we find for all 21 BCG Mg$_{2}$ gradients. The correlation for the SB06 elliptical sample (with velocity dispersion larger than 2.2 km s$^{-1}$) are shown with the dashed grey line, and the correlation for the Carollo et al.\ (1993) sample are shown with the dot-dash grey line.}
   \label{fig:Mg2Grads}
\end{figure}

We plot the Mg$_{2}$ gradients against central velocity dispersion in Figure \ref{fig:Mg2Grads}, using different symbols to indicate the gradients of the galaxies where the slit was placed within 15 degrees of the MA (where the MA was known), and the non-significant gradients. We do a linear fit inversely weighted by the errors on both the Mg$_{2}$ gradients as well as the velocity dispersion for all the BCGs, regardless of slit orientation, and find a correlation with a slope of --0.060 $\pm$ 0.049. We find a probability of 23 per cent ($P$=0.229) that these two parameters are not related according to a statistical t-test at 95 per cent confidence level. 

When we fit a relation to the nine galaxies for which we have derived gradients along the major axis, we find a slope of --0.085 $\pm$ 0.067 between these two parameters, and a probability of 25 per cent ($P$=0.248) that these two parameters are not related according to a statistical t-test at 95 per cent confidence level. When we do the fit to the 17 significant gradients, we find a slope of --0.072 $\pm$ 0.047, and a probability of 15 per cent ($P$=0.146). It can be seen that, contrary to the previous findings for elliptical galaxies not in the centre of clusters, the gradients seem to become steeper as the velocity dispersion increases although, admittedly, the result is not very significant and will need to be confirmed with larger data sets.

We note that this behaviour can also be seen in Brough et al.\ (2007), where they analyse the metallicity gradients of six BCGs and brightest group galaxies. The Mg$_2$ gradients against velocity dispersion of three out of their five galaxies that falls in our BCG mass range show the same steep Mg$_{2}$ gradients (also see Spolaor et al.\ 2009, where they show that the SSP metallicity gradients for three BCGs from the Brough et al.\ (2007) sample show steeper metallicity gradients than ellipticals of the same mass). 

For a concrete comparison with normal high-mass ellipticals, we also investigate the Mg$_{2}$ gradients against sigma for ellipticals with $\log \sigma>$ 2.2 km s$^{-1}$ in the S\'{a}nchez-Bl\'{a}zquez et al.\ (2006a, hereafter SB06) and Carollo et al.\ (1993) studies. We use 39 galaxies from SB06 and 26 from Carollo et al.\ (1993), regardless of slit orientation, to do linear fits inversely weighted by the errors on both the Mg$_{2}$ gradients and the velocity dispersions. We find non-significant relations (probabilities $P$=0.904 and $P$=0.846, and slopes --0.008 $\pm$ 0.068 and 0.013 $\pm$ 0.063 respectively), also shown in Figure \ref{fig:Mg2Grads}. The normal massive ellipticals show flat correlations (the slope of the correlation with velocity dispersion for both elliptical samples is consistent with zero), and the correlations are much more scattered than the one for the BCGs (where the slope is --0.060 $\pm$ 0.049).

We also performed K--S tests on the distributions of the Mg$_{2}$ gradients (non-significant gradients included) of the three samples against each other, where the null hypothesis is that the distributions were drawn from an identical parent population. The critical value of the statistical test, at a 95 per cent confidence level, is $D=1.36\sqrt{\frac{m+n}{m \times n}}$, where $m$ and $n$ are the number of galaxies in the sample. Thus, if the test value is larger than $D$, then the samples compared are significantly different from each other at the 95 per cent confidence level. We find that the distribution of the BCG sample is significantly different from the Carollo et al.\ (1993) sample at the 95 per cent confidence level (test value 0.570, compared to $D=0.403$), but consistent with the distribution of the SB06 sample (test value 0.215, compared to $D=0.370$). We also find that the distributions of the two elliptical samples (Carollo et al.\ (1993) and SB06) are significantly different from each other (test value 0.488, compared to $D=0.350$), emphasizing the large scatter generally observed in the Mg$_{2}$ gradients of massive ellipticals. It should also be emphasized that K--S tests are related to the mean values and dispersions and not to the slopes of the correlations.

We also plot the Mg$_{2}$ gradients against the central Mg$_{2}$ measurements in Figure \ref{fig:Mg2Grads}, and find an absence of a correlation. Thus, the steepening of gradient with mass is not due to the sole increasing metallicity of the galactic core. Many authors have however detected such a correlation in their samples of elliptical galaxies (Gonz\'{a}lez $\&$ Gorgas 1995; S\'{a}nchez-Bl\'{a}zquez et al.\ 2006b; Kuntschner et al.\ 2006).

In dissipative collapse models, the gas flows to the centre of the galaxy, and is continually enriched by evolving stars (Eggen, Lynden-Bell $\&$ Sandage 1962; Carlberg 1984; Arimoto $\&$ Yoshii 1987; Gibson 1997). The depth of the galaxy potential well regulates the efficiency of the dissipational processes, so that a strong correlation between metallicity gradient and galaxy mass is expected in this case (Kawata $\&$ Gibson 2003). Therefore, and as discussed in Paper 1, formation by dissipative collapse at $z > 2$ predicts strong negative metallicity gradients that correlate with galaxy mass. 

On the other hand, according to the De Lucia $\&$ Blaizot (2007) simulations, we expect to see evidence of dissipationless mergers in these galaxies. Signatures then include shallow metallicity gradients and little dependence of metallicity on mass (Bekki $\&$ Shioya 1999; Kobayashi 2004). From the predictions by De Lucia $\&$ Blaizot (2007), more massive galaxies in high density environments are expected to have shallower negative metallicity gradients (the metallicity gradients in BCG are therefore expected to be related to the properties of the host cluster).

\subsection{BCG offset from the cluster X-ray peak}
\label{cluster}

The Mg$_{2}$ gradient -- velocity dispersion correlation for BCGs, that is stronger than that of elliptical galaxies in the same mass range, is possibly a consequence of the special location of BCGs in the centre of the cluster gravitational well. Hierarchical models of galaxy formation predict that the formation of the central galaxy is closely connected with the evolution of the host cluster. In principle, this would not necessarily be reflected by differences in the star formation histories (and therefore the indices), as stars might have formed before the formation -- or assembly -- of the real galaxy. However, it is interesting to investigate whether, and to what extent, the characteristics of the host cluster play a role in the properties of BCGs.

\begin{figure}
   \centering
   \mbox{\subfigure{\includegraphics[scale=0.65]{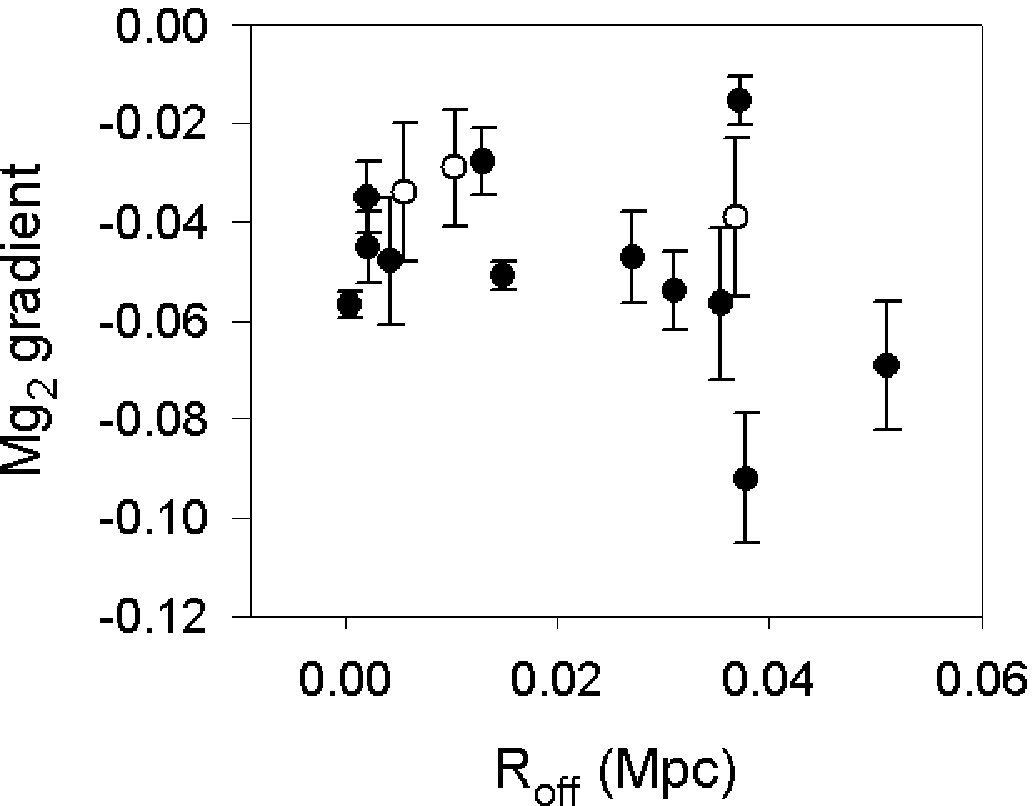}}}
         \mbox{\subfigure{\includegraphics[scale=0.65]{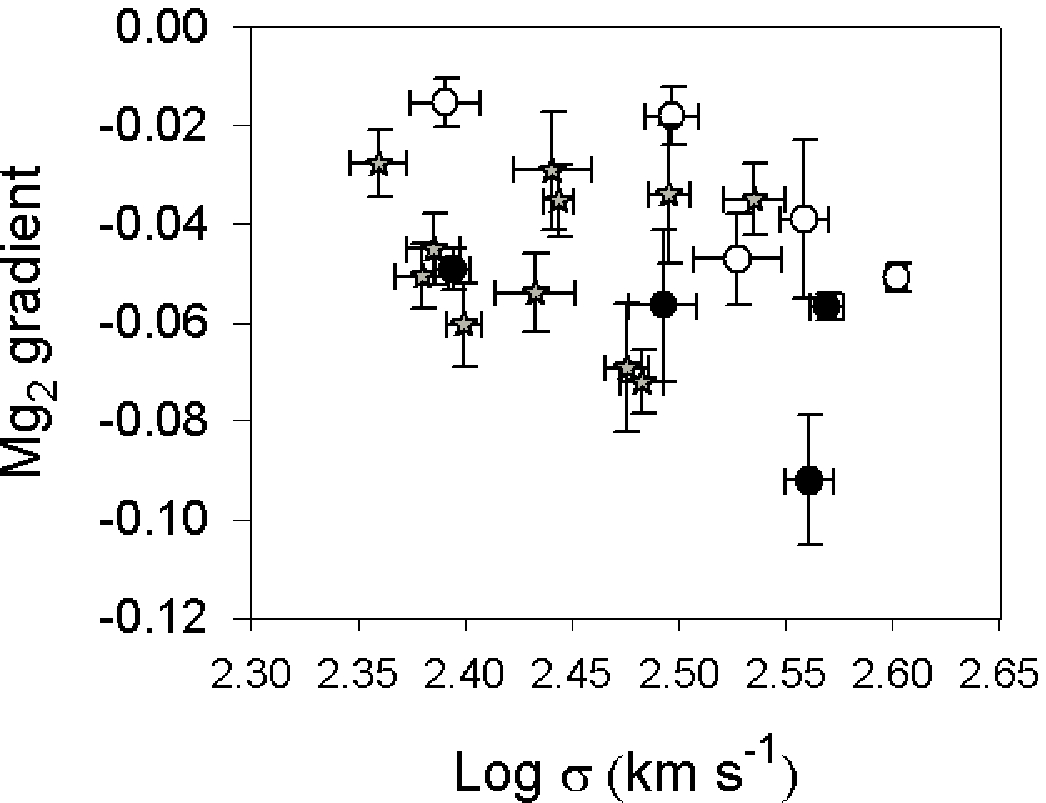}}}
   \caption{The upper plot shows the Mg$_{2}$ gradients against projected distance between BCG and X-ray peak (the empty symbols indicate the non-significant gradients). The lower plot shows the Mg$_{2}$ gradients (non-significant gradients included) against BCG velocity dispersion where different symbols indicate whether or not the BCG is hosted by a cooling flow cluster (the grey stars indicate the non-cooling flow clusters, the empty circles indicate the cooling flow clusters, and the black filled circles indicate the clusters where the cooling flow status is not known).}
   \label{fig:ClusterProp}
\end{figure}

Numerical simulations predict that the offset of the BCG from the peak of the cluster X-ray emission is an indication of how close the cluster is to the dynamical equilibrium state, and that this decreases as the cluster evolves (Katayama et al.\ 2003). We plot the Mg$_{2}$ gradients against the X-ray offset $R_{\rm off}$ (from Table \ref{table:objects}) in Figure \ref{fig:ClusterProp}. The X-ray offset $R_{\rm off}$ was calculated using $H_{0}$ = 75 km s$^{-1}$ Mpc$^{-1}$ in Mpc. We find a significant correlation between Mg$_{2}$ gradients and $R_{\rm off}$ with a probability of $P=0.044$ (therefore the probability that these two parameters are not related is four per cent according to a statistical t-test at 95 per cent confidence level) when all the Mg$_{2}$ gradients are included, and an even stronger correlation ($P<0.0001$) when only the significant gradients are included.

We also correlate the Mg$_{2}$ gradients against four other host cluster properties (X-ray luminosity; X-ray temperature; cluster velocity dispersion; and whether or not the host cluster is a cooling flow cluster or not), but no other statistically significant correlations were found. The plot indicating the cooling flow status of the host cluster is also shown in Figure \ref{fig:ClusterProp} (obtained from Table 7 in Paper 2 and Table A1 in Paper 3).

\section{Conclusions}
\label{conclusions}

We find, for a sample of 21 BCGs with Mg$_{2}$ gradients (17 of which significant) a weak correlation between the Mg$_{2}$ gradients and central velocity dispersion, such that BCG Mg$_{2}$ gradients become steeper with increasing mass. For normal ellipticals, in the same mass range, we find correlations consistent with zero slopes. In general, metallicity gradients are found to be flat or decrease with increasing mass for ordinary massive elliptical galaxies (and with a large scatter). We notice that this steepening of BCG gradients can also be seen in Spolaor et al.\ (2009, including the six BCGs from Brough et al.\ (2007), their figure 1 - red and black triangles) where they show the metallicity gradients\footnote{We use Mg$_{2}$ instead of metallicity gradients here.} of eight BCGs and brightest group galaxies. Five out of their eight galaxies show steeper metallicity gradients than ellipticals of the same mass. The steep negative metallicity gradients that correlate with mass is not reproduced by simulations even for massive ordinary elliptical galaxies (Hopkins et al.\ 2009; Pipino et al.\ 2010). The result of the correlation between the Mg$_2$ gradient and velocity dispersion is, admittedly, not very significant, and it will need to be confirmed with larger data sets.  

However, we also find a strong correlation between BCG distance to the X-ray peak and Mg$_{2}$ gradients. This correlation indicates a relation between the gradients and the dynamical state of the cluster, and probably a correlation between the way the mass was assembled in the BCG and the dynamical state of the cluster. Thus, these correlations may give us information about the formation and evolution of the clusters themselves.

\section*{Acknowledgments}
We thank the anonymous referee for constructive comments which contributed to the improvement of this paper. PSB is supported by the Ministerio de Ciencia e Innovaci\'on (MICINN) of Spain through the  Ramon y Cajal programme. This work has been supported by the Programa Nacional de Astronom\'{\i}a y Astrof\'{\i}sica of the Spanish Ministry of Science and Innovation under the grant AYA2007-67752-C03-01. Based on observations obtained on the Gemini North and South telescopes.

\bsp

\label{lastpage}

\end{document}